\documentclass[12pt]{article}
\usepackage{latexsym,epsfig,graphicx,amsmath,amssymb,amscd}
\usepackage{multirow,paralist,dsfont,url}
\usepackage{slashbox}
\usepackage[titletoc]{appendix}

\usepackage[american]{babel}
\usepackage{xcolor}
\usepackage{grffile}
\usepackage{xcolor}

\usepackage{natbib}  %\citet{xxx}, (\citealt{xxx})
\usepackage{enumitem}
\usepackage{comment}

\newcommand{\thetavec}{{\boldsymbol{\theta}}}

\newcommand{\betahat}{{\widehat{\beta}}}
\newcommand{\Ahat}{{\widehat{A}}}
\newcommand{\phat}{{\widehat{p}}}
\newcommand{\degreeC}{{\,\ensuremath{^\circ}\text{C}}}

\newcommand{\NOR}{{\rm N}}
\newcommand{\muhat}{\widehat{\mu}}
\newcommand{\sigmahat}{\widehat{\sigma}}

\newcommand{\thetavechat}{\widehat{\thetavec}}

\newcommand{\cminv}{\,\textrm{cm}^{-1}}

\newcommand{\xvec}{\boldsymbol{x}}

\newcommand{\tran}{^\top}

\newcommand{\nm}{{\,\text{nm}}}
\newcommand{\pect}{{\,\text{\%}}}
\newcommand{\TempC}{\text{TempC}}
\newcommand{\TempK}{\text{TempK}}
\newcommand{\RH}{\text{RH}}
\newcommand{\temp}{\text{temp}}
\newcommand{\rh}{\text{rh}}
\newcommand{\xtemp}{\texttt{xtemp}}
\newcommand{\xrh}{\texttt{xrh}}
\newcommand{\loglik}{\mathcal{L}}
\newcommand{\approxsim}{{\,\dot{\sim}\,}}
\newcommand{\new}{\text{new}}

\newtheorem{prealgo}{Algorithm}

\newenvironment{algorithm}[1]%
{\begin{prealgo}\upshape \textbf{#1.}}{\end{prealgo}}

\begin{document}

%%%%%%%%%%%%TITLE%%%%%%%%%%%%%%%%%%%%%%%%%%%%%%%%%%%
\title{Statistical and Deep Learning Approaches for Predicting Degradation of Polymeric Materials in Photovoltaics}
%%%%%%%%%%%%%%%%%%%%%%%%%%%%%%%%%%%%%%%%%%%%%%%%%%%%%%%%%%%%%%%%%%%%%%%%%%%%%%%%%%%%%%%%%%%%%%%%%%%%%%%%%%%%%%%%%

\author{
Yili Hong$^1$\footnote{Corresponding Author. Email: yilihong@vt.edu}\,\, and Xiaohong Gu$^2$\\[1.5ex]
{\small $^1$Department of Statistics, Virginia Tech, Blacksburg, VA 24061}\\
{\small $^2$Engineering Laboratory, National Institute of Standards and Technology,}\\[-0.75ex]
{\small Gaithersburg, MD 20899}
}

%\date{\today}
\date{}

\maketitle
%%%%%%%%%%%%%%%%%%%%%%%%%%%%%%%%%%%%%%%%%%%%%%%%%%%%%%%%%%%%%%%%%%%%%%%%%%%%%%%%%%%%%%%%%%%%%%%%%%%%%%%%%%%%%%%%
\begin{abstract}
Polymeric materials are widely used in photovoltaic (PV) systems, making it essential to understand their service life to ensure reliable PV performance. The primary failure mechanism of polymeric materials in PV systems is photodegradation caused by ultraviolet (UV) radiation. Degradation modeling provides a framework for predicting service life, with a key step being the development of predictive models for degradation paths. This paper presents statistical and machine learning approaches for predicting the outdoor degradation of polymeric components in PV systems. We describe the study design and data collection process for developing predictive models based on indoor laboratory testing data, which are then extended to outdoor field conditions with time-varying environmental variables, with prediction uncertainty quantified through simulation. Deep learning (DL) methods are also explored, and results are compared across modeling approaches. The parametric statistical model demonstrates good fit and predictive performance across datasets and shows greater robustness by incorporating physical and chemical knowledge. The DL model provides flexibility in capturing complex covariate relationships and often yields accurate predictions, though it is less robust across datasets. The paper concludes with remarks on key findings and their implications for PV reliability.

\end{abstract}

\newpage
%\tableofcontents
%\newpage

%%%%%%%%%%%%%%%%%%%%%%%%%%%%%%%%%%%%%%%%%%%%%%%%%%%%%%%%%%%%%%%%%%%%%%%%%%%%%%%%%%%%%%%
\section{Introduction}
%%%%%%%%%%%%%%%%%%%%%%%%%%%%%%%%%%%%%%%%%%%%%%%%%%%%%%%%%%%%%%%%%%%%%%%%%%%%%%%%%%%%%%%
\subsection{Background and Motivation}
%%%%%%%%%%%%%%%%%%%%%%%%%%%%%%%%%%%%%%%%%%%%%%%%%%%%%%%%%%%%%%%%%%%%%%%%%%%%%%%%%%%%%%%
The service life prediction of photovoltaic (PV) systems is a critical area of study for ensuring their long-term deployment. PV systems contain numerous polymeric components, including encapsulants, frontsheets, backsheets, edge sealants, and junction boxes. These components and their interfaces are susceptible to degradation and loss of functional properties under harsh environmental conditions. Therefore, predicting the degradation of these polymeric materials is a key aspect of the broader goal of PV system service life prediction. This paper focuses specifically on the degradation prediction of these polymeric materials.

The first step in this process is to collect accelerated degradation test data under controlled indoor conditions. Using this data, together with knowledge of the underlying physics and chemistry of the degradation process, one can identify appropriate functional forms that describe how experimental variables relate to the degradation path. These functional forms, along with the accelerated test data, are then used to construct a degradation path model that links specimen degradation trajectories to the experimental variables. The resulting predictive model can be applied to forecast degradation for specified covariate histories. To assess the effectiveness of the accelerated testing methodology, predictions from the model, based on accelerated test data, are compared against the observed degradation paths of specimens exposed to outdoor environments.

To implement the above strategy, both indoor and outdoor test data are required. The indoor test data are obtained under controlled laboratory conditions, while the outdoor test data are collected from specimens exposed to real-world environments. At the
National Institute of Standards and Technology (NIST),
accelerated laboratory tests have been conducted using the SPHERE (Simulated Photodegradation via High-Energy Radiant Exposure) facility to investigate the effects of weathering variables on the degradation of polymeric materials. A commercial multilayered PV backsheet, polyethylene terephthalate (PET)/PET/ ethylene-vinyl acetate (EVA), referred to as PPE,  is used as the model in these accelerated tests. Those tests provide data to quantify the influence of environmental factors, such as light intensity and wavelength, temperature, and relative humidity, on material degradation. Of particular interest are the degradation rates indicated by the growth of the yellowness index (YI) and other chemical changes, which are found to be closely linked to environmental factors including UV spectrum, UV intensity, temperature, and relative humidity~(RH). The outdoor experiments are conducted at three locations under four different exposure settings: Arizona, Florida, Maryland rack (on outdoor rack), and Maryland box. For the Maryland box setting, the bottom of the box chamber was made of black-anodized aluminum, the top was covered with borofloat glass, and all sides were enclosed with a breathable fabric material (\citealt{Guetal2008}).

However, quantitative relationships for degradation prediction must first be established from indoor test data using statistical models and then verified by linking them to outdoor exposure results. Once developed from indoor accelerated testing data and appropriately connected to outdoor conditions, these models can be used to predict the degradation of samples exposed outdoors. The predictions can then be compared with observed outdoor degradation data to validate the modeling approach. In addition to statistical modeling and prediction, machine learning techniques, particularly deep learning (DL), can also be applied to degradation prediction. In this paper, we further investigate the use of DL methods to predict the degradation levels of polymeric materials in PV systems.

In summary, the objective of this paper is to develop statistical and machine learning methods to predict the degradation of polymeric components in photovoltaics. These methods can be further applied to investigate service life and the complex property-performance relationships of engineering polymers used in PV encapsulation under varying climatic conditions.

%%%%%%%%%%%%%%%%%%%%%%%%%%%%%%%%%%%%%%%%%%%%%%%%%%%%%%%%%%%%%%%%%%%%%%%%%%%%%%%%%%%%%%%
\subsection{Related Literature and Contributions of This Work}
%%%%%%%%%%%%%%%%%%%%%%%%%%%%%%%%%%%%%%%%%%%%%%%%%%%%%%%%%%%%%%%%%%%%%%%%%%%%%%%%%%%%%%%
In this section, we review the literature in the following areas: service life and degradation modeling of polymeric materials used in PV systems and related organic coatings, statistical methods for service life modeling and prediction, and general literature on DL and its implementation.

In the literature on the reliability and durability of PV materials, \citet{Gambogi2013Weathering} highlighted that polymeric packaging plays a critical role in ensuring the reliability of PV modules. In particular, the backsheet serves as the first line of defense against environmental weathering factors. PPE is a widely used backsheet material due to its low cost (\citealt{Lin2016DepthProfiling}), which highlights the need to study its degradation behavior. \citet{Guetal2017PVSC} examined the impact of UV light intensity on the photodegradation of PPE backsheets. \citet*{lyu2018service_life_pv} quantified the effects of light intensity and wavelength on the discoloration of a glass/EVA/PPE laminate. \citet{Fairbrotheretal2018SolarEnergy} analyzed differential degradation patterns of PV backsheets at the array level for non-fluoropolymer-based backsheets. \citet{Guetal2020PVSC} developed a methodology for predicting the long-term performance of PPE backsheets. In addition to the literature on PPE, \citet{ZielnikBurns2018} discussed the durability and reliability of polymers and other materials used in photovoltaic modules. \citet{Lyuetal2018JPV} investigated the cracking behavior of polyamide-based backsheets under sequential fragmentation testing. \citet{Lyuetal2020} examined how environmental variables influence the degradation of photovoltaic components, particularly for polyamide-based backsheets.

In the area of service life prediction for polymeric materials, relevant literature can be found on coating service life prediction. \citet{MartinLechnerVarner1994} provided a quantitative characterization of photodegradation effects in polymeric materials exposed to weathering environments. \citet{Rabek1995} is a valuable resource on polymer photodegradation mechanisms and experimental methods. \citet{Martin1996} discussed methodologies for predicting the service life of coating systems, while \citet{MartinChinNguyen2003} reviewed reciprocity law experiments in polymeric photodegradation. \citet{Chinetal2004} developed an accelerated UV weathering device based on integrating sphere technology. \citet{Guetal2008} demonstrated how accelerated laboratory tests can be linked to outdoor performance results for a model epoxy coating system.

In the area of statistical modeling for degradation data from polymeric materials, \citet{VacaTrigoMeeker2009} proposed a statistical model to link field and laboratory exposure results for a model coating. \citet{HongDuanetal2015} modeled the effects of time-varying covariates on the outdoor degradation of coatings, while \citet{Duanetal2017} used indoor test data to predict outdoor degradation for coatings. For DL and its implementation, \citet{Goodfellow-et-al-2016} provides a comprehensive resource on DL concepts. Deep learning models can be implemented using programming frameworks such as PyTorch (\citealt{PyTorch_NEURIPS2019_9015}).  \citet{Songetal2024ELimage} applied DL techniques to detect defects in electroluminescence (EL) images of solar panels. \citet{clark2025quality} provided practitioners with a comprehensive introduction to degradation models.

Drawing from our literature review, we identified a notable gap: the absence of case studies on predicting the degradation of specific polymeric materials, such as the PPE system used in PV applications. To address this, our research makes several key contributions. First, we introduce a unique, large-scale dataset of PPE degradation collected over three years from both controlled indoor experiments and real-world outdoor tests. Using this dataset, we develop a statistical predictive model with meaningful physical interpretation, capable of generating outdoor predictions with quantified uncertainty. We then validate the underlying degradation mechanisms by comparing model predictions with outdoor test data. In parallel, we develop a physics-informed DL model and assess its effectiveness in degradation prediction. Finally, we provide practical guidance on how to use the PPE dataset, together with the statistical and DL approaches, to study the degradation of a broader class of materials.

%%%%%%%%%%%%%%%%%%%%%%%%%%%%%%%%%%%%%%%%%%%%%%%%%%%%%%%%%%%%%%%%%%%%%%%%%%%%%%%%%%%%%%%
\subsection{Overview}
%%%%%%%%%%%%%%%%%%%%%%%%%%%%%%%%%%%%%%%%%%%%%%%%%%%%%%%%%%%%%%%%%%%%%%%%%%%%%%%%%%%%%%%

The remainder of the paper is organized as follows. Section~\ref{sec:study.design.data.collection} outlines the study design and data collection for polymeric materials in PV systems. Section~\ref{sec:statistical.approach} presents the statistical approach for building the predictive model, including model fitting and prediction.
Section~\ref{sec:indoor.data.analysis} presents the analysis of the indoor data and the interpretation of the results. Section~\ref{sec:outdoor.prediction} describes the prediction of outdoor degradation based on environmental data and reports the corresponding results. Section~\ref{sec:deep.learning} introduces a DL model for outdoor predictions. Finally, Section~\ref{sec:conclusion} summarizes the findings and suggests directions for future research.

%%%%%%%%%%%%%%%%%%%%%%%%%%%%%%%%%%%%%%%%%%%%%%%%%%%%%%%%%%%%%%%%%%%%%%%%%%%%%%%%%%%%%%%
\section{Study Design and Data Collection}\label{sec:study.design.data.collection}
%%%%%%%%%%%%%%%%%%%%%%%%%%%%%%%%%%%%%%%%%%%%%%%%%%%%%%%%%%%%%%%%%%%%%%%%%%%%%%%%%%%%%%%
\subsection{Study Design}
%%%%%%%%%%%%%%%%%%%%%%%%%%%%%%%%%%%%%%%%%%%%%%%%%%%%%%%%%%%%%%%%%%%%%%%%%%%%%%%%%%%%%%%

We first introduce the data collected at the
NIST Engineering Laboratory
for building a predictive model, using PPE as the material of interest. Two degradation indices are considered: discoloration and chemical change.  Discoloration, also referred to as the yellowness index, is measured with a colorimeter. A higher yellowness index value indicates a greater degree of yellowing. Chemical changes are quantified as the ratio of two chemical compounds, specifically the peaks at 1245$\cminv$ and 1410$\cminv$ in the FTIR spectrum. These peaks correspond to two chemical structures of particular interest in the study. In particular, 1245$\cminv$ corresponds to the asymmetric C-C-O stretching vibration, while 1410$\cminv$ corresponds to the in-plane vibrations of the aromatic ring. The damage data (i.e., yellowness index and chemical change) were obtained from the measurement of exposed PET outer layer side. For further details on these measurements, we refer the reader to \citet{Guetal2020PVSC}.

We focus on discoloration and chemical change in this paper because they capture complementary aspects of degradation: discoloration reflects physical appearance, while chemical change tracks underlying reactions. They also exhibit opposite trends (increasing vs. decreasing), providing a useful illustration for modeling. The proposed method, however, is general and applicable to other degradation metrics.

\begin{figure}
\begin{center}
\includegraphics[width=.48\textwidth]{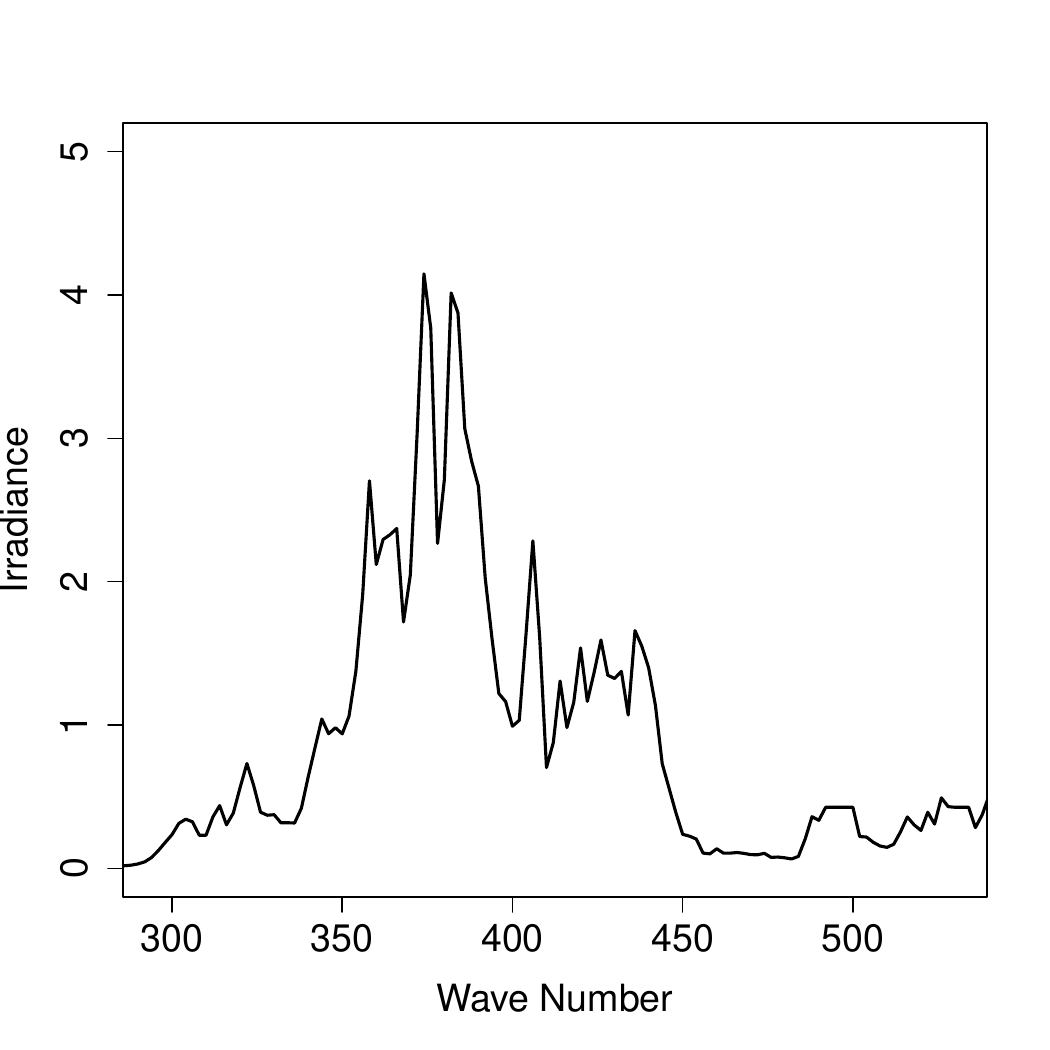}
\caption{Illustrations of the irradiance from the lamp used in the PPE indoor experiments. }\label{fig:irrad.plot}
\end{center}
\end{figure}

\begin{figure}
\begin{center}
\begin{tabular}{cc}
\includegraphics[width=.48\textwidth]{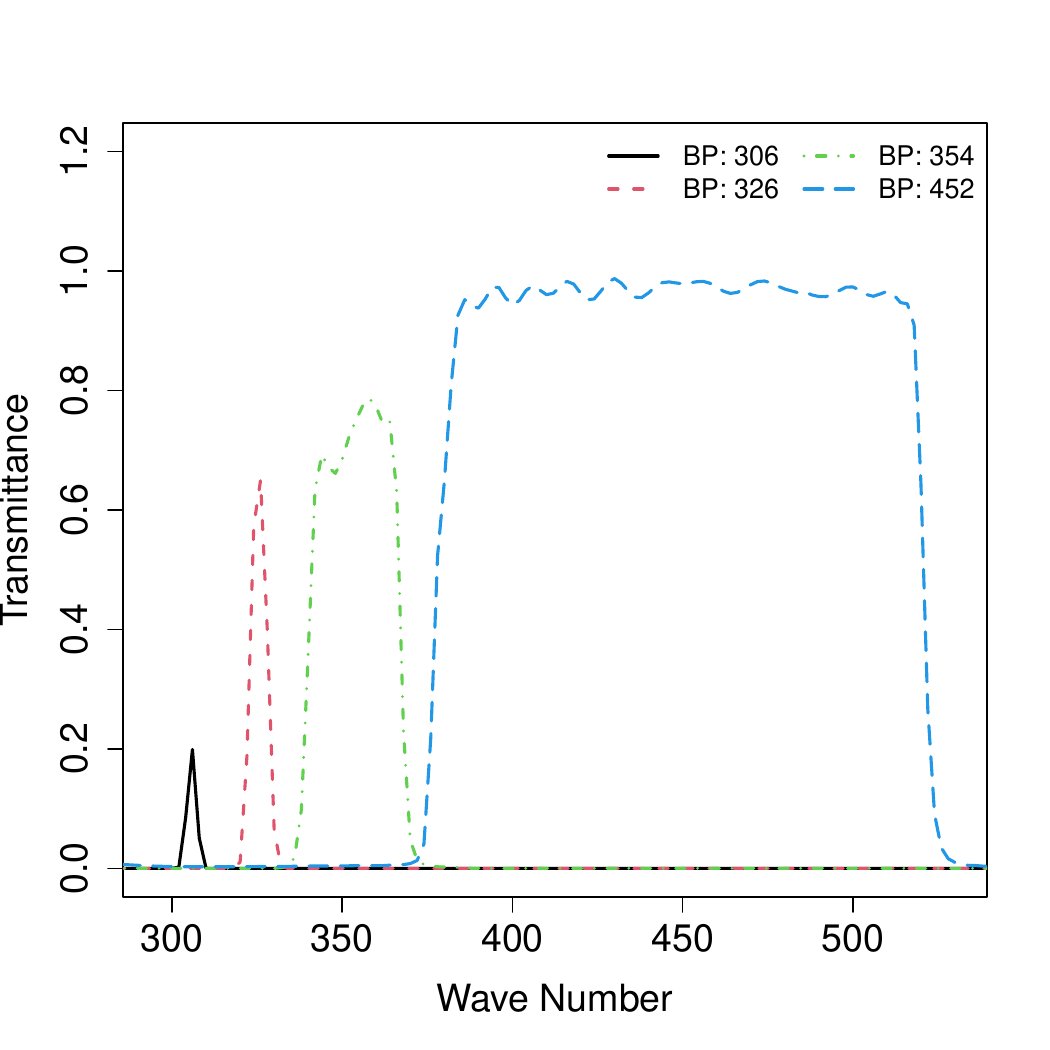}&
\includegraphics[width=.48\textwidth]{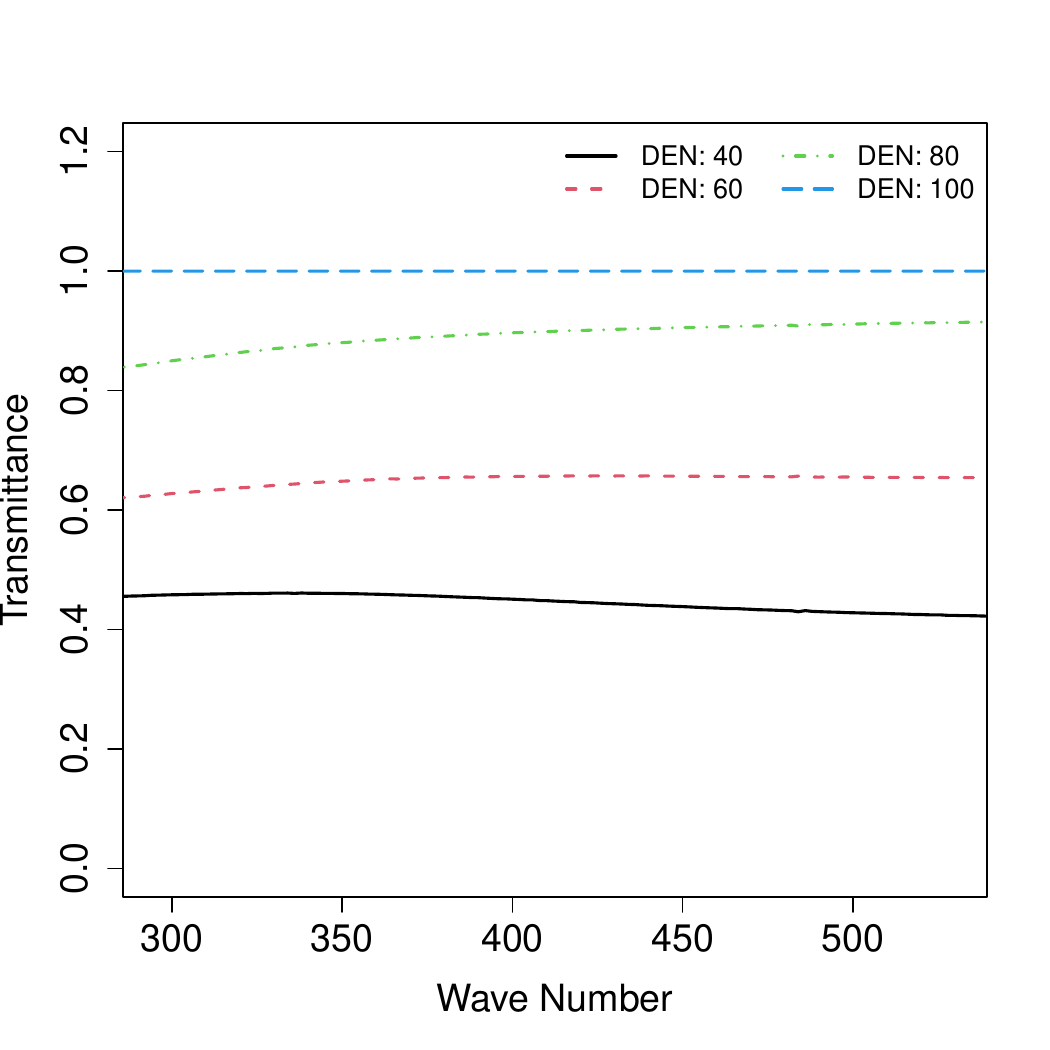}\\
(a) Band-pass Filters  & (b) Neutral Density Filters
\end{tabular}
\caption{Illustrations of band-pass and neutral density filters used in the PPE indoor experiments. Keys: BP = bandpass filter, DEN = neutral density filter.
}\label{fig:filters.plot}
\end{center}
\end{figure}

The degradation of PPE in PV systems is primarily driven by UV exposure. In the indoor experiments, the light source is a lamp (see Figure~\ref{fig:irrad.plot}). For modeling, we denote the spectral irradiance of this lamp by $E_i(\lambda)$, where $\lambda$ is the wavelength and $i$ is the index for specimen. For UV radiation, two factors are particularly important: the spectrum and the intensity. Shorter wavelengths cause more damage than longer wavelengths, and higher intensity (i.e., more photons) accelerates degradation. Temperature and relative humidity also influence the process. Accordingly, the experimental factors in this study include UV spectrum, UV intensity, temperature, and relative humidity (RH).

The effect of the UV spectrum (i.e., wavelength effect) is examined using band-pass (BP) filters, which selectively transmit light within a specified range of wavelengths. Four BP filters were employed, with center wavelengths and bandwidths of 306$\nm$ ($\pm$3$\nm$), 326$\nm$ ($\pm$6$\nm$), 353$\nm$ ($\pm$19$\nm$), and 452$\nm$ ($\pm$80$\nm$). For instance, the 306$\nm$ BP filter transmits light centered at 306$\nm$ within a range of $\pm$3$\nm$. Figure~\ref{fig:filters.plot}(a) illustrates the four band-pass filters used in the PPE indoor experiments.

The effect of UV intensity is investigated using neutral density (ND) filters, which reduce the overall light intensity without significantly altering the spectral shape. Four ND filters were employed, with nominal transmittance levels of 40$\pect$, 60$\pect$, 80$\pect$, and 100$\pect$. For instance, a 40$\pect$ ND filter transmits 40$\pect$ of the light emitted by the lamp. Figure~\ref{fig:filters.plot}(b) illustrates the four neutral density filters used in the PPE indoor experiments.

For each specimen, the incident light first passes through the band-pass filter, followed by the neutral density filter. The combined transmittance of these filters is represented by $F_i(\lambda)$ for specimen $i$, expressed as a function of wavelength $\lambda$. In the statistical analysis, we focus on the portion of the spectrum relevant to photodegradation by setting transmittance values to zero for wavelengths above 550$\nm$. This truncation reflects that longer wavelengths do not contribute meaningfully to the degradation process and can therefore be excluded from the modeling framework.

The temperature levels considered in the experiments were 45$\degreeC$, 65$\degreeC$, 75$\degreeC$, and 85$\degreeC$, while the RH levels were set at 0$\pect$ and 60$\pect$. In total, seven experiments were conducted, as summarized in the first column of Table~\ref{tab:experiment.setup.YI}. The reciprocity law study, denoted by ``\_R'' in Table~\ref{tab:experiment.setup.YI}, investigates whether material degradation depends solely on the cumulative light dose rather than the irradiance intensity. For further details on the reciprocity law, we refer the reader to~\citet{MartinChinNguyen2003}.
The spectral study, indicated by ``\_W,'' investigates the effect of the UV spectrum. It is expected that shorter wavelengths (e.g., around 306$\nm$) cause much greater damage to materials than longer wavelengths (e.g., around 452$\nm$).

In practical applications, the solar spectral distribution varies with location and time, making it difficult to reproduce the full UV spectrum using laboratory light sources. However, the wavelength and reciprocity studies allow one to relate indoor and outdoor exposure results, despite differences in the light sources.

\begin{table}[h]
\caption{Summary of indoor experimental setups and samples measured for the yellowness index. Samples highlighted in black are used to train the statistical and machine learning models, while samples highlighted in red are used to evaluate their performance. Keys: EID = experiment ID, T = temperature, RH = relative humidity, BP = bandpass filter, DEN = neutral density filter, SID = sample ID.}\label{tab:experiment.setup.YI}
\begin{center}
\begin{small}
\begin{tabular}{cccccc|cccccc}\hline\hline
EID	&	T	&	RH	&	BP	&	DEN	&	SID	&	EID	&	T	&	RH	&	BP	&	DEN	&	SID	\\\hline
45\_0\_R	&	45	&	0	&	-	&	40$\pect$	&	8, 11, 14, 17	&	75\_0\_W	&	75	&	0	&	-	&	100$\pect$	&	2, 3, \textcolor{red}{4, 5}	\\
	&	45	&	0	&	-	&	60$\pect$	&	7, 10, 13, 16	&		&	75	&	0	&	306	&	-	&	8, 12, 16	\\
	&	45	&	0	&	-	&	80$\pect$	&	6, 9, 12, 15	&		&	75	&	0	&	326	&	-	&	9, 13, 17	\\
	&	45	&	0	&	-	&	100$\pect$	&	2, \textcolor{red}{3,} 4, \textcolor{red}{5}	&		&	75	&	0	&	354	&	-	&	6, 10, 14	\\
	&		&		&		&		&		&		&	75	&	0	&	389	&	-	&	7, 11, 15	\\\hline
65\_0\_R	&	65	&	0	&	-	&	40$\pect$	&	2, 8, 12, 16	&	85\_0\_W	&	85	&	0	&	-	&	100$\pect$	&	\textcolor{red}{2,} 3, 4,\textcolor{red}{ 5}	\\
	&	65	&	0	&	-	&	60$\pect$	&	3, 9, 13, 17	&		&	85	&	0	&	306	&	-	&	8, 12, 16	\\
	&	65	&	0	&	-	&	80$\pect$	&	4, 6, 10, 14	&		&	85	&	0	&	326	&	-	&	9, 13, 17	\\
	&	65	&	0	&	-	&	100$\pect$	&	5, \textcolor{red}{7, 11}, 15	&		&	85	&	0	&	354	&	-	&	6, 10, 14	\\
	&		&		&		&		&		&		&		85&		0&	389	&	-	&	7, 11, 15	\\\hline
65\_0\_W	&	65	&	0	&	-	&	100$\pect$	&	\textcolor{red}{2, 3}, 4, 5	&	85\_60\_R	&	85	&	60	&	-	&	40$\pect$	&	2, 8, 12, 16	\\
	&	65	&	0	&	306	&	-	&	8, 12, 16	&		&	85	&	60	&	-	&	60$\pect$	&	3, 9, 13, 17	\\
	&	65	&	0	&	326	&	-	&	9, 13, 17	&		&	85	&	60	&	-	&	80$\pect$	&	4, 6, 10, 14	\\
	&	65	&	0	&	354	&	-	&	6, 10, 14	&		&	85	&	60	&	-	&	100$\pect$	&	5, \textcolor{red}{7, 11}, 15	\\
	&	65	&	0	&	389	&	-	&	7, 11, 15	&		&		&		&		&		&		\\\hline
75\_0\_R	&	75	&	0	&	-	&	40$\pect$	&	8, 11, 14, 17	&		&		&		&		&		&		\\
	&	75	&	0	&	-	&	60$\pect$	&	7, 10, 13, 16	&		&		&		&		&		&		\\
	&	75	&	0	&	-	&	80$\pect$	&	6, 9, 12, 15	&		&		&		&		&		&		\\
	&	75	&	0	&	-	&	100$\pect$	&	2, \textcolor{red}{3, 4}, 5	&		&		&		&		&		&		\\\hline
\hline
\end{tabular}
\end{small}
\end{center}
\end{table}

%%%%%%%%%%%%%%%%%%%%%%%%%%%%%%%%%%%%%%%%%%%%%%%%%%%%%%%%%%%%%%%%%%%%%%%%%%%%%%%%%%%%%%%
\subsection{Indoor Data}
%%%%%%%%%%%%%%%%%%%%%%%%%%%%%%%%%%%%%%%%%%%%%%%%%%%%%%%%%%%%%%%%%%%%%%%%%%%%%%%%%%%%%%%

For the indoor laboratory test data, referred to as the indoor dataset, a total of 112 samples were studied across seven experiments, with 16 samples in each experiment. As shown in Table~\ref{tab:experiment.setup.YI}, about 3 to 4 samples have been exposed at each experimental condition. All samples were measured for yellowness index, while only a subset was measured for chemical change. The allocation of samples with chemical change measurements is provided in Table~\ref{tab:CR.sample.list} and was predetermined prior to the study, with a total of 62 samples.

Note that the experimental plan shown in Table~\ref{tab:experiment.setup.YI} is not a full factorial design. Implementing a full factorial design is impractical due to time and budget constraints. Instead, a hybrid design approach was adopted, integrating domain knowledge of the material with statistical design principles to enable efficient data collection. Although the design is not fully factorial, the resulting data remain well suited for developing predictive degradation models by combining statistical modeling with underlying physical and chemical insights.

\begin{figure}%[h]
\begin{center}
\begin{tabular}{cc}
\includegraphics[width=.48\textwidth]{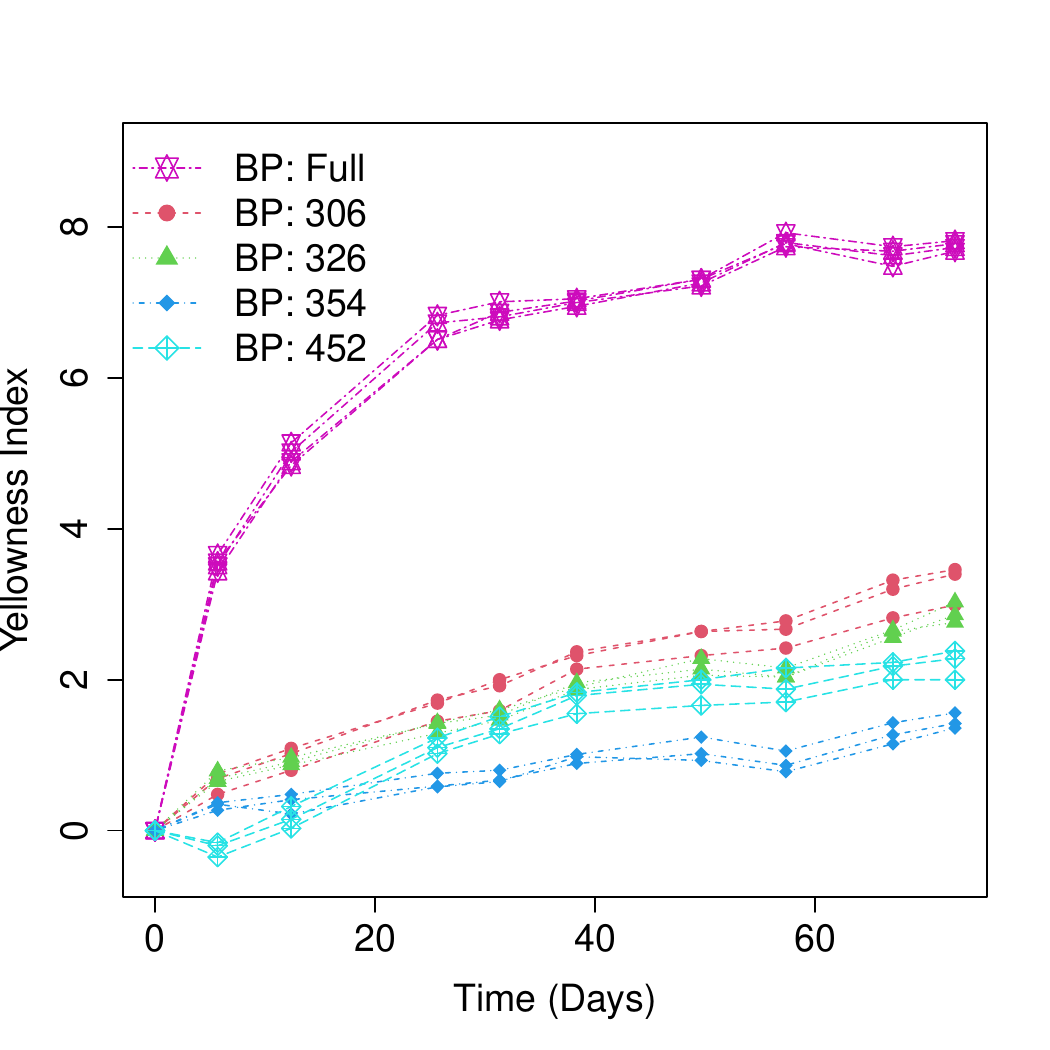}&
\includegraphics[width=.48\textwidth]{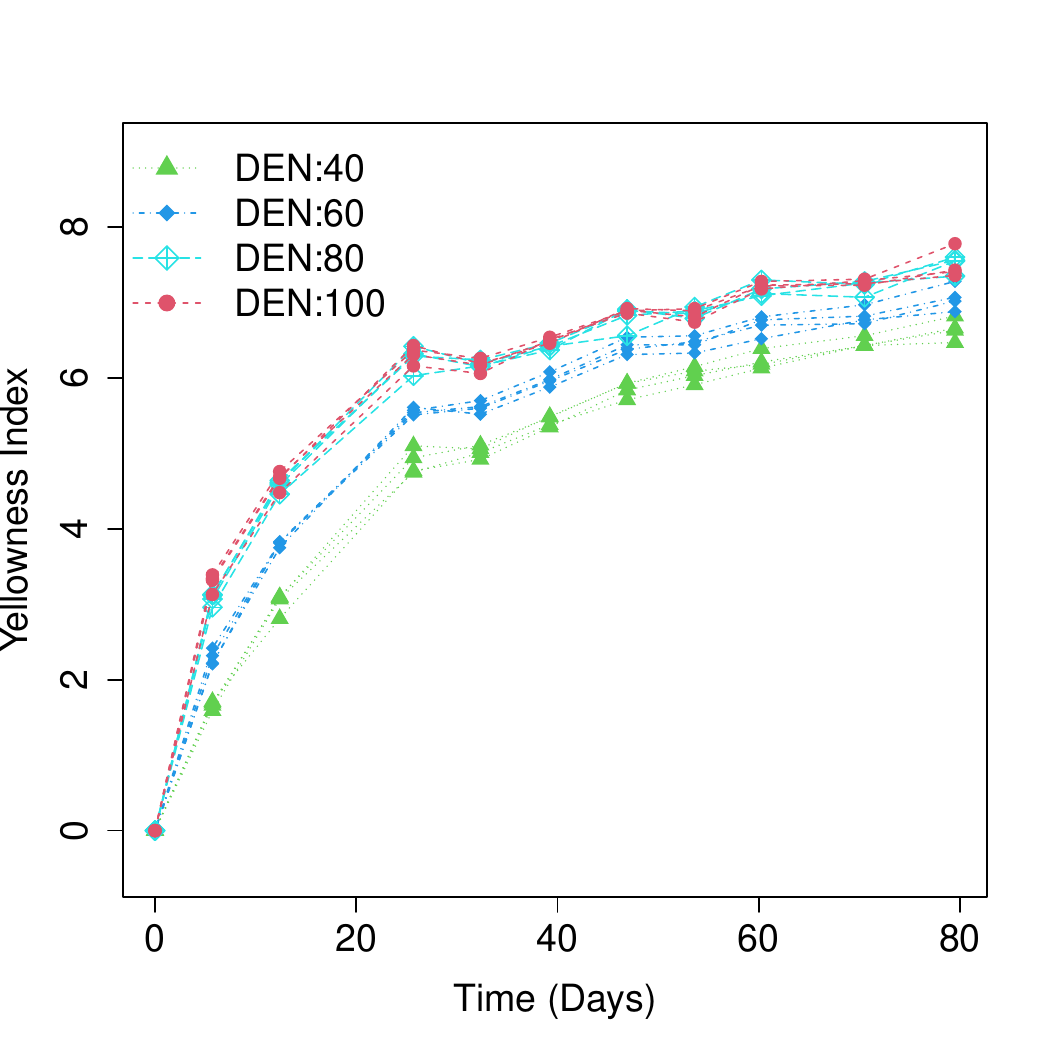}\\
(a) Wavelength Study (EID: 85\_0\_W) & (b) Reciprocity Study (EID: 75\_0\_R)
\end{tabular}
\caption{Visualization of the yellowness index as a function of time from a subset of the indoor data.}\label{fig:YI.indoor.data.plot}
\end{center}
\end{figure}

\begin{figure}[h]
\begin{center}
\begin{tabular}{cc}
\includegraphics[width=.48\textwidth]{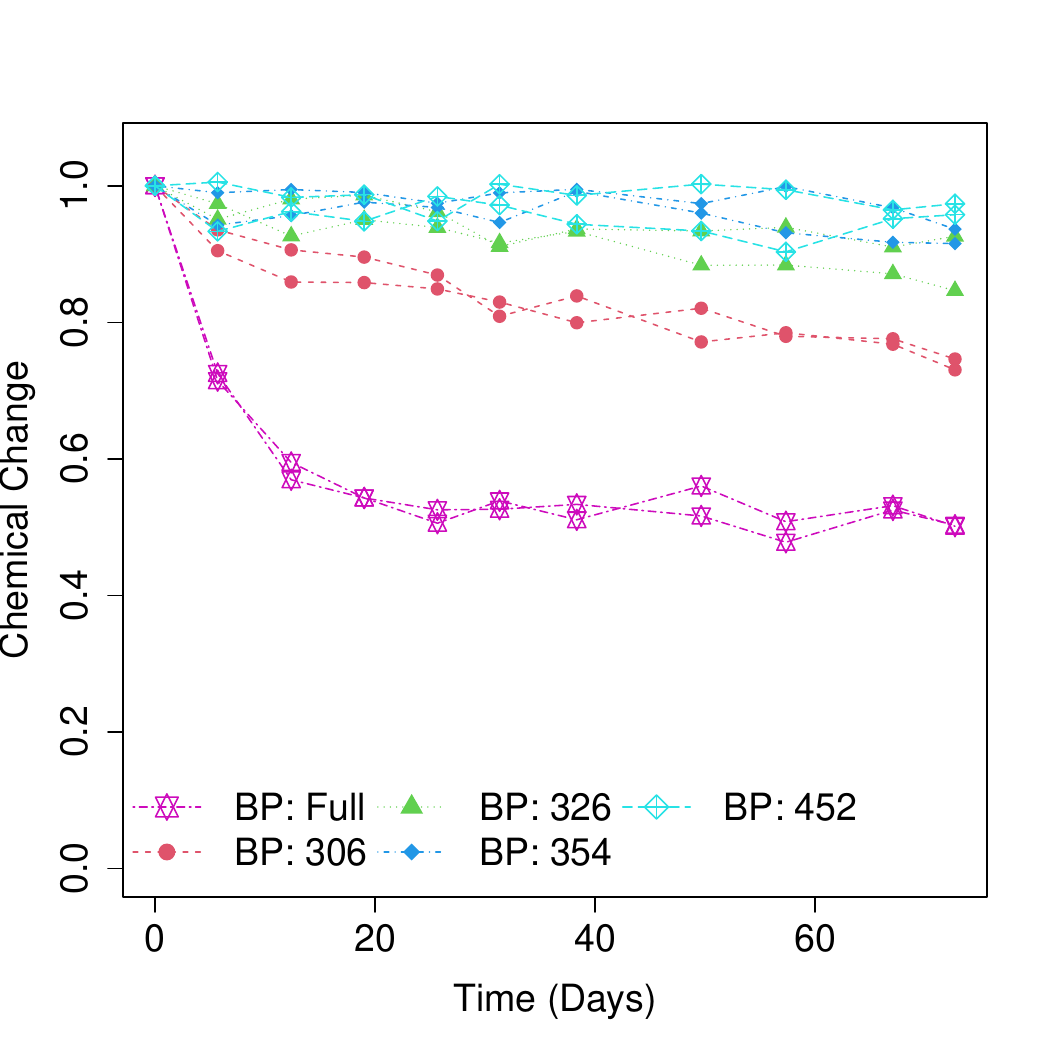}&
\includegraphics[width=.48\textwidth]{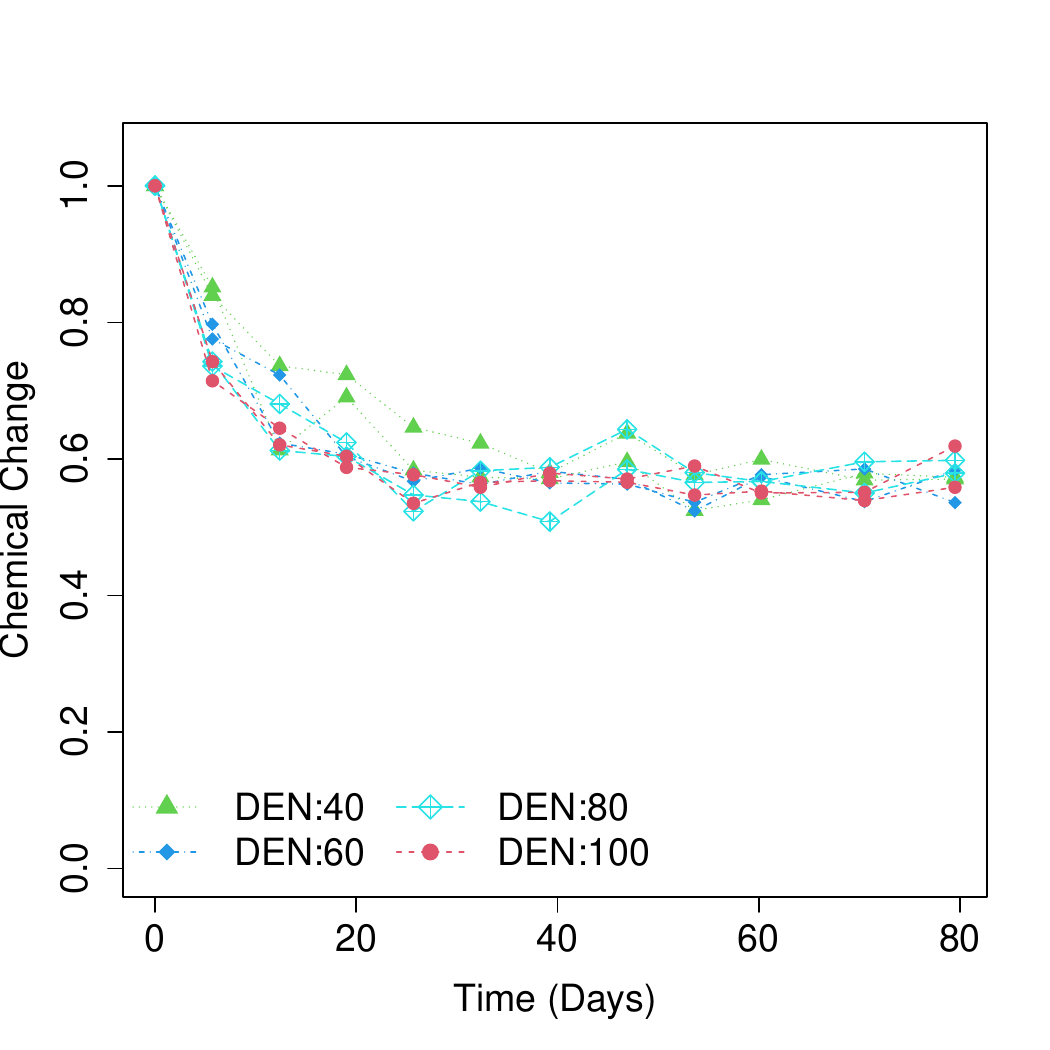}\\
(a) Wavelength Study (EID: 85\_0\_W)  & (b) Reciprocity Study (EID: 75\_0\_R)
\end{tabular}
\caption{Visualization of the chemical change (i.e., the ratio of two chemical compounds, characterized by peaks at 1245$\cminv$ and 1410$\cminv$ in the FTIR spectrum
) as a function of time from a subset of the indoor data. }\label{fig:CR.indoor.data.plot}
\end{center}
\end{figure}

Figure~\ref{fig:YI.indoor.data.plot} presents the measurements of yellowness index as a function of time for a subset of the indoor data, where time is measured in days since the start of the experiment. Panel (a) shows degradation data from a wavelength study (experiment ID: 85\_0\_W). The degradation paths exhibit an initial increase followed by a plateau. The full-wavelength setting shows the greatest degradation, as it receives the highest UV radiation from the lamp. Samples under the BP: 306\,nm setting also degrade rapidly, despite the small dose transmitted through the filter, highlighting the strong effect of shorter UV wavelengths. Panel (b) displays data from a reciprocity study (experiment ID: 75\_0\_R). Here, the effect of UV intensity is less pronounced, as the degradation paths overlap in some cases. Figure~\ref{fig:CR.indoor.data.plot} presents analogous results for chemical change, showing similar overall patterns, except that the degradation trend is decreasing over time.

For the purpose of building and validating predictive models, the set of samples was divided into training and test sets for both statistical and machine learning analyses, as summarized in Tables~\ref{tab:experiment.setup.YI} and~\ref{tab:CR.sample.list}. The training set, shown in black, was used to fit the models and estimate parameters, ensuring that the models could capture the underlying degradation patterns under a variety of experimental conditions. The test set, highlighted in red, was reserved exclusively for performance evaluation, providing an assessment of the models' predictive accuracy. For the yellowness index analysis, the training and test sets contain 98 and 14 samples, respectively (Table~\ref{tab:experiment.setup.YI}; total $=112$), while for the chemical change analysis, the corresponding sample sizes are 55 and 7 (Table~\ref{tab:CR.sample.list}; total $=62$).

We further clarify the rationale behind the training–test split. Because our objective differs from a typical machine learning setting, the split departs from standard practice. Specifically, our goal is not to predict all indoor experimental conditions, but to develop a model capable of predicting degradation under outdoor conditions, which most closely reflect the real application scenario. The outdoor test samples (introduced in Section~\ref{sec:out.data.intro}) were obtained from materials exposed to natural environmental conditions for three years, providing a realistic test scenario. Given the limited number of outdoor samples, we use as much data as possible for model training to ensure robust estimation, while reserving a subset of indoor samples for additional testing. For the indoor test set, we select samples exposed under the 100\% transmittance and full-wavelength condition, as these exhibit the greatest degradation during the experimental period. Other exposure conditions generally lead to less severe degradation, making the 100\% intensity and full-wavelength condition the most challenging scenario for prediction; thus, it serves as a stringent test of model performance.

\begin{table}[h]
\caption{List of samples measured for the chemical change. Samples highlighted in black are used to train the statistical and machine learning models, while samples highlighted in red are used to evaluate their performance. }\label{tab:CR.sample.list}
\begin{center}
\begin{small}
\begin{tabular}{cccccc|cccccc}\hline\hline
EID	&	T	&	RH	&	BP	&	DEN	&	SID	&	EID	&	T	&	RH	&	BP	&	DEN	&	SID	\\\hline
45\_0\_R	&	45	&	0	&	-	&	40$\pect$	&	8, 11	&	75\_0\_W	&	75	&	0	&	-	&	100$\pect$	&	2, \textcolor{red}{3}	\\
	&	45	&	0	&	-	&	60$\pect$	&	7, 10	&		&	75	&	0	&	306	&	-	&	8, 12	\\
	&	45	&	0	&	-	&	80$\pect$	&	6, 9	&		&	75	&	0	&	326	&	-	&	9, 13	\\
	&	45	&	0	&	-	&	100$\pect$	&	2, \textcolor{red}{3}	&		&	75	&	0	&	354	&	-	&	6, 10 	\\
	&		&		&		&		&		&		&	75	&	0	&	452	&	-	&	7, 15	\\\hline
65\_0\_R	&	65	&	0	&	-	&	40$\pect$	&	2, 8	&	85\_0\_W	&	85	&	0	&	-	&	100$\pect$	&	\textcolor{red}{2}, 3	\\
	&	65	&	0	&	-	&	60$\pect$	&	3, 9	&		&	85	&	0	&	306	&	-	&	8, 12	\\
	&	65	&	0	&	-	&	80$\pect$	&	4, 6 	&		&	85	&	0	&	326	&	-	&	9, 13	\\
	&	65	&	0	&	-	&	100$\pect$	&	\textcolor{red}{5}, 7	&		&	85	&	0	&	354	&	-	&	6, 10 	\\
	&		&		&		&		&		&		&		85&		0&	452	&	-	&	7, 15	\\\hline
65\_0\_W	&	65	&	0	&	-	&	100$\pect$	&	2, \textcolor{red}{3}	&	85\_60\_R	&	85	&	60	&	-	&	40$\pect$	&	2, 8	\\
	&	65	&	0	&	306	&	-	&	8, 12	&		&	85	&	60	&	-	&	60$\pect$	&	3, 9	\\
	&	65	&	0	&	326	&	-	&	9, 13	&		&	85	&	60	&	-	&	80$\pect$	&	4, 6 	\\
	&	65	&	0	&	354	&	-	&	6, 10 	&		&	85	&	60	&	-	&	100$\pect$	&	\textcolor{red}{5}, 7	\\
	&	65	&	0	&	452	&	-	&	7, 15	&		&		&		&		&		&		\\\hline
75\_0\_R	&	75	&	0	&	-	&	40$\pect$	&	8, 11	&		&		&		&		&		&		\\
	&	75	&	0	&	-	&	60$\pect$	&	7, 10	&		&		&		&		&		&		\\
	&	75	&	0	&	-	&	80$\pect$	&	6, 9	&		&		&		&		&		&		\\
	&	75	&	0	&	-	&	100$\pect$	&	\textcolor{red}{2}, 3	&		&		&		&		&		&		\\\hline
\hline
\end{tabular}
\end{small}
\end{center}
\end{table}

%%%%%%%%%%%%%%%%%%%%%%%%%%%%%%%%%%%%%%%%%%%%%%%%%%%%%%%%%%%%%%%%%%%%%%%%%%%%%%%%%%%%%%%
\subsection{Outdoor Data}\label{sec:out.data.intro}
%%%%%%%%%%%%%%%%%%%%%%%%%%%%%%%%%%%%%%%%%%%%%%%%%%%%%%%%%%%%%%%%%%%%%%%%%%%%%%%%%%%%%%%

As mentioned earlier, the outdoor experiments were conducted at three locations with four exposure settings: Arizona (AZ), Florida (FL), Maryland (MD) rack, and Maryland (MD) box. In both AZ and FL, the specimens were mounted on racks facing the sun. At the MD site, two configurations were used: one with specimens placed on a rack and the other with specimens enclosed in a glass box, resulting in four distinct exposure settings. The outdoor experiment lasted for more than two years.

\begin{figure}
\begin{center}
\begin{tabular}{cc}
\includegraphics[width=.48\textwidth]{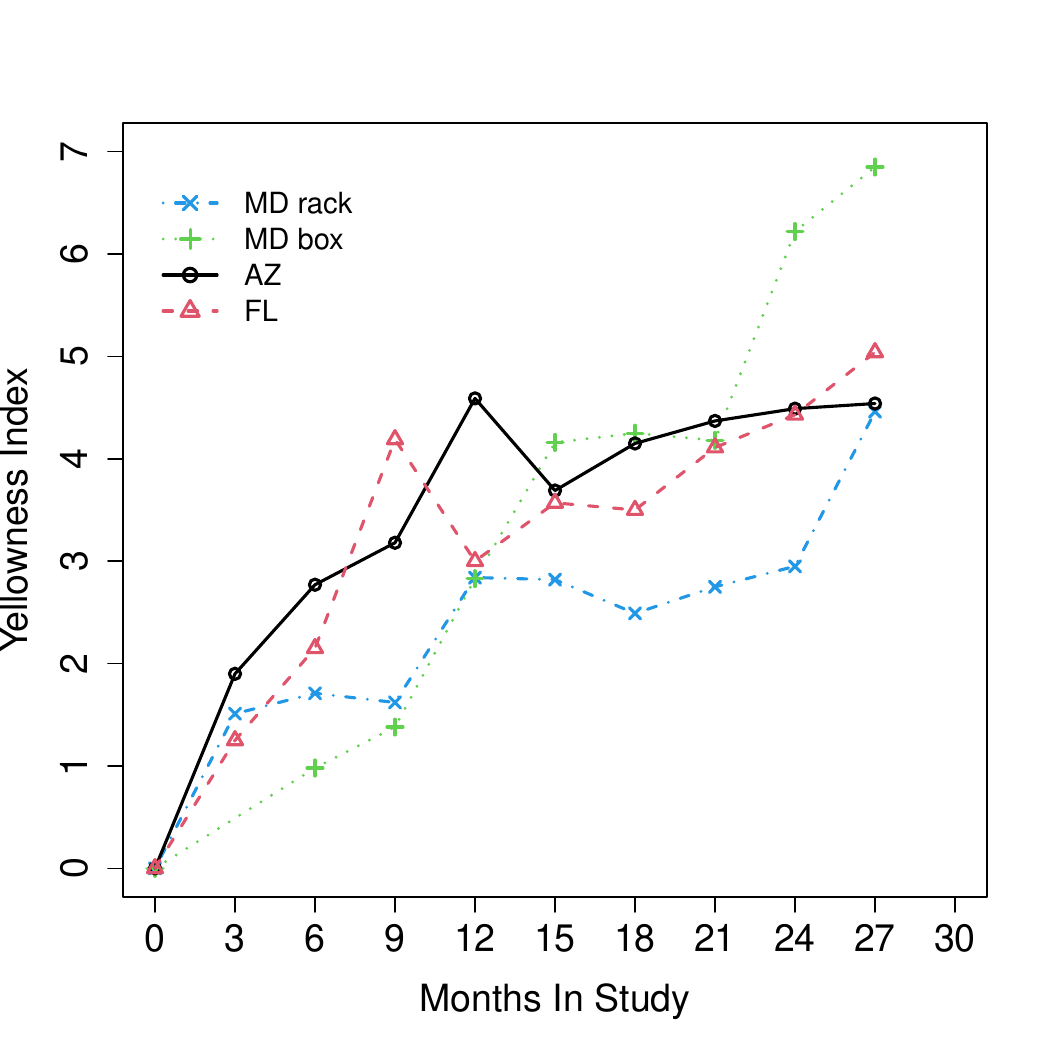}&
\includegraphics[width=.48\textwidth]{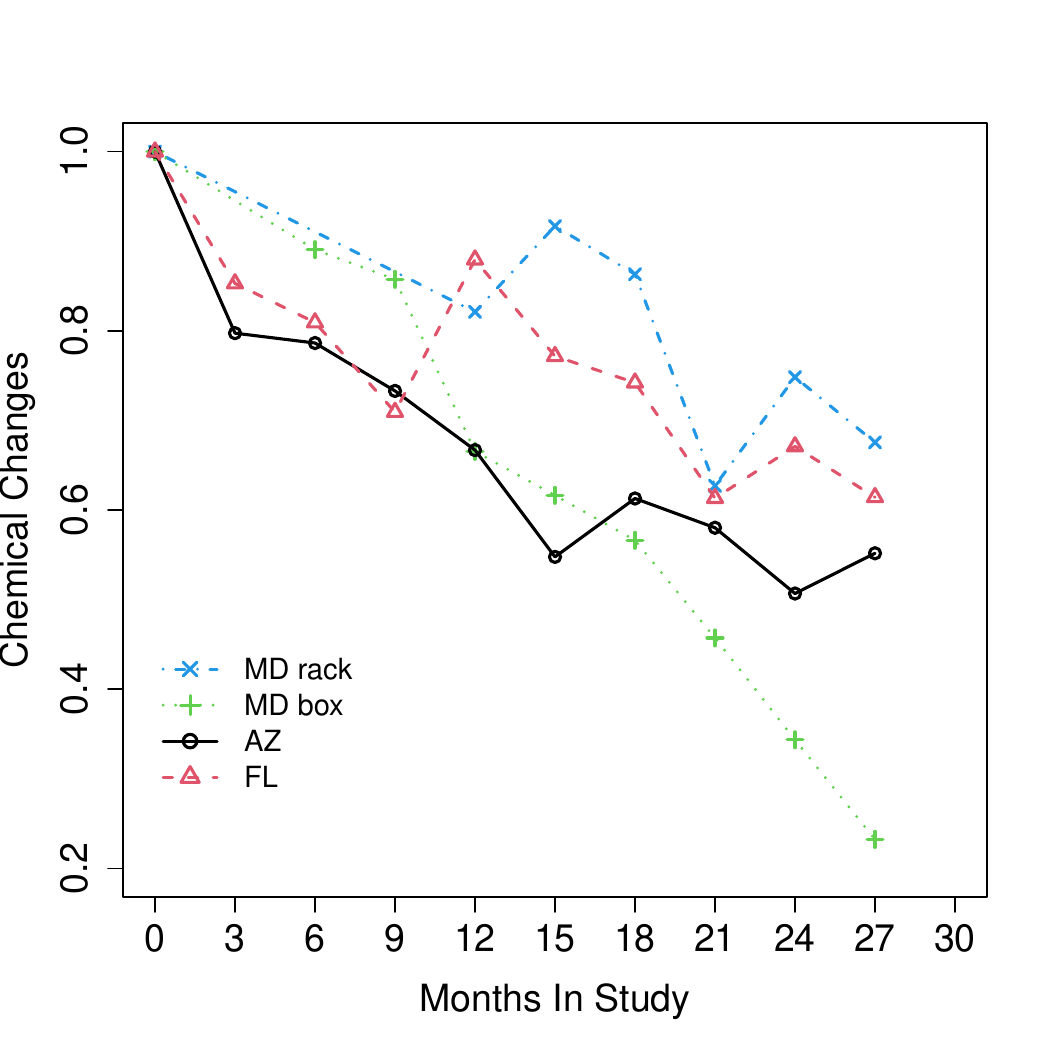}\\
(a) Yellowness Index & (b) Chemical Change
\end{tabular}
\caption{Plot of outdoor degradation measurements for yellowness index~(a) and chemical change~(b) as functions of time. The $x$-axis indicates the number of months since July 2017.
}\label{fig:outdoor.deg.data.plot}
\end{center}
\end{figure}

Measurements of the outdoor units were taken every three months, during which a small piece of each sample was cut and sent to the laboratory for analysis. Figure~\ref{fig:outdoor.deg.data.plot} presents the outdoor degradation measurements for yellowness index and chemical change as functions of time, with the $x$-axis indicating the number of months since July 2017. For YI, the general trend is increasing, whereas for chemical change, the trend is decreasing. The outdoor degradation data also exhibit greater variability compared to the indoor data.

For the outdoor data, time-varying environmental covariates were also recorded, including solar irradiance, temperature, and relative humidity (RH), with hourly resolution. The weather data span the period from 2017 to 2019, corresponding to the duration of the outdoor tests. For solar irradiance, an irradiance profile, a curve as a function of wavelength $\lambda$, was recorded at each time point $\tau$ and is denoted by $E_i(\lambda, \tau)$. Temperature and RH were recorded as time series, denoted by $\TempC_i(\tau)$ and $\RH_i(\tau)$, respectively.

\begin{figure}
\begin{center}
\begin{tabular}{cc}
\includegraphics[width=.48\textwidth]{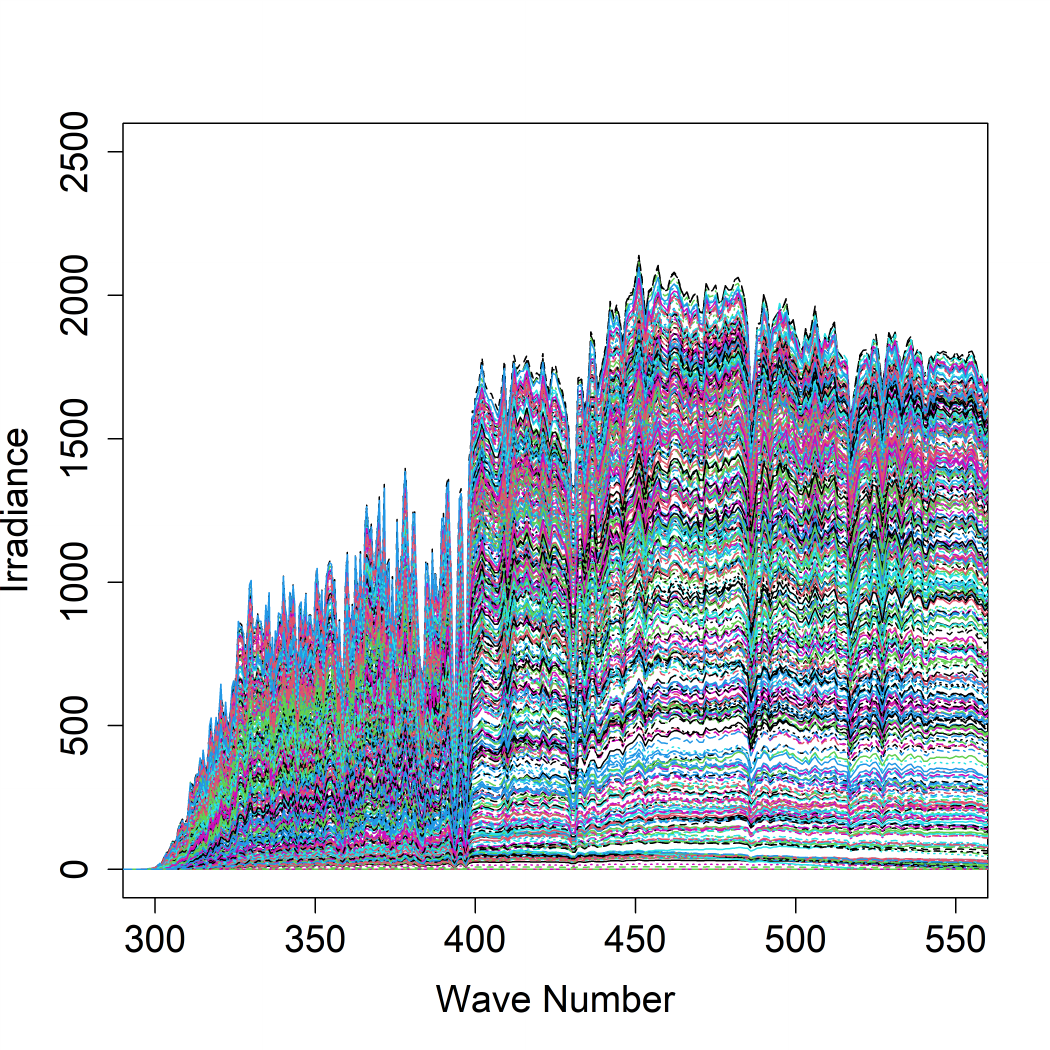}&
\includegraphics[width=.48\textwidth]{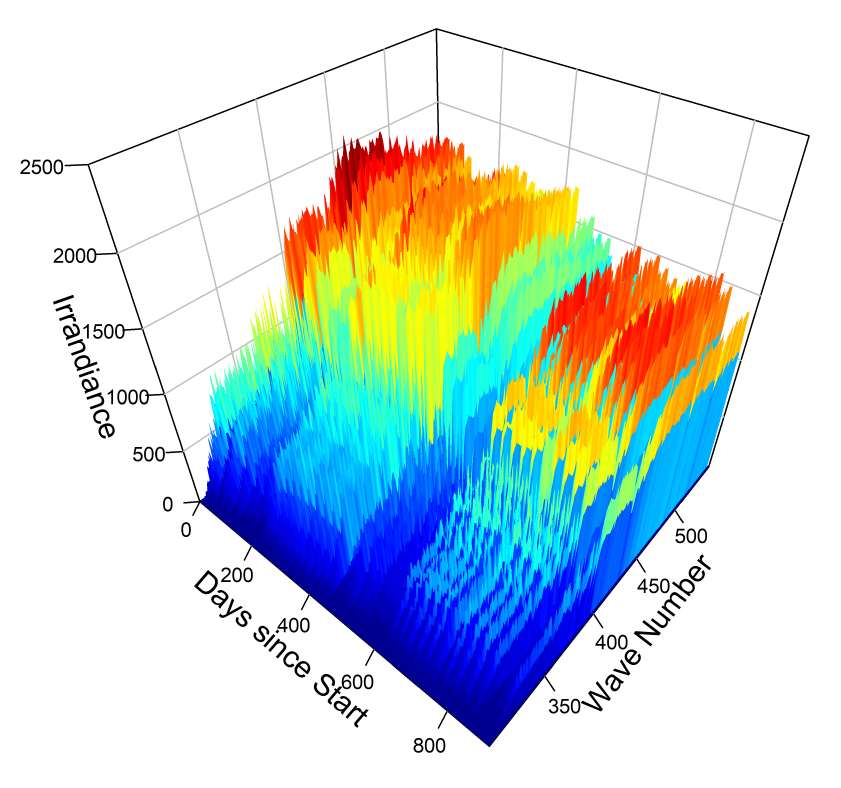}\\
(a) Irradiance Profile & (b) Irradiance 3D\\
\includegraphics[width=.48\textwidth]{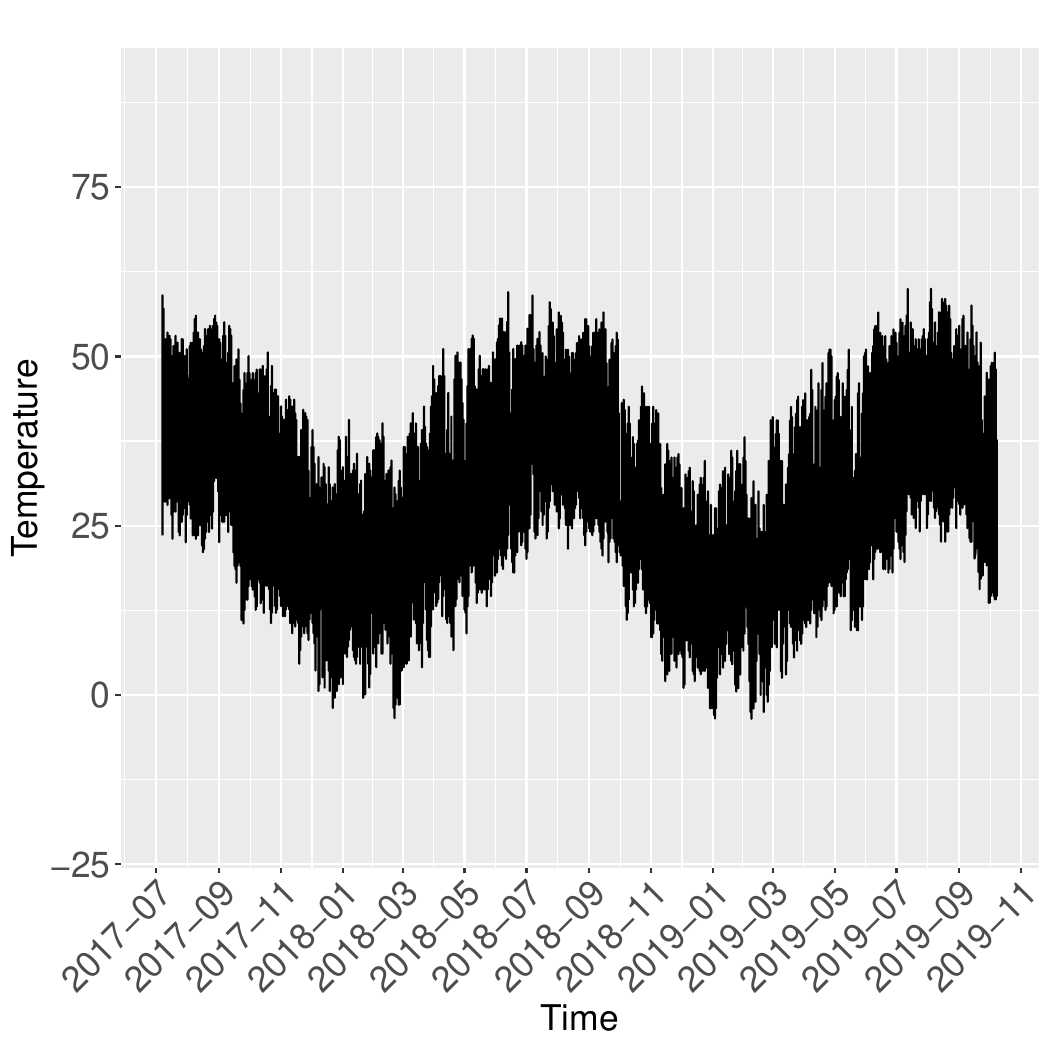}&
\includegraphics[width=.48\textwidth]{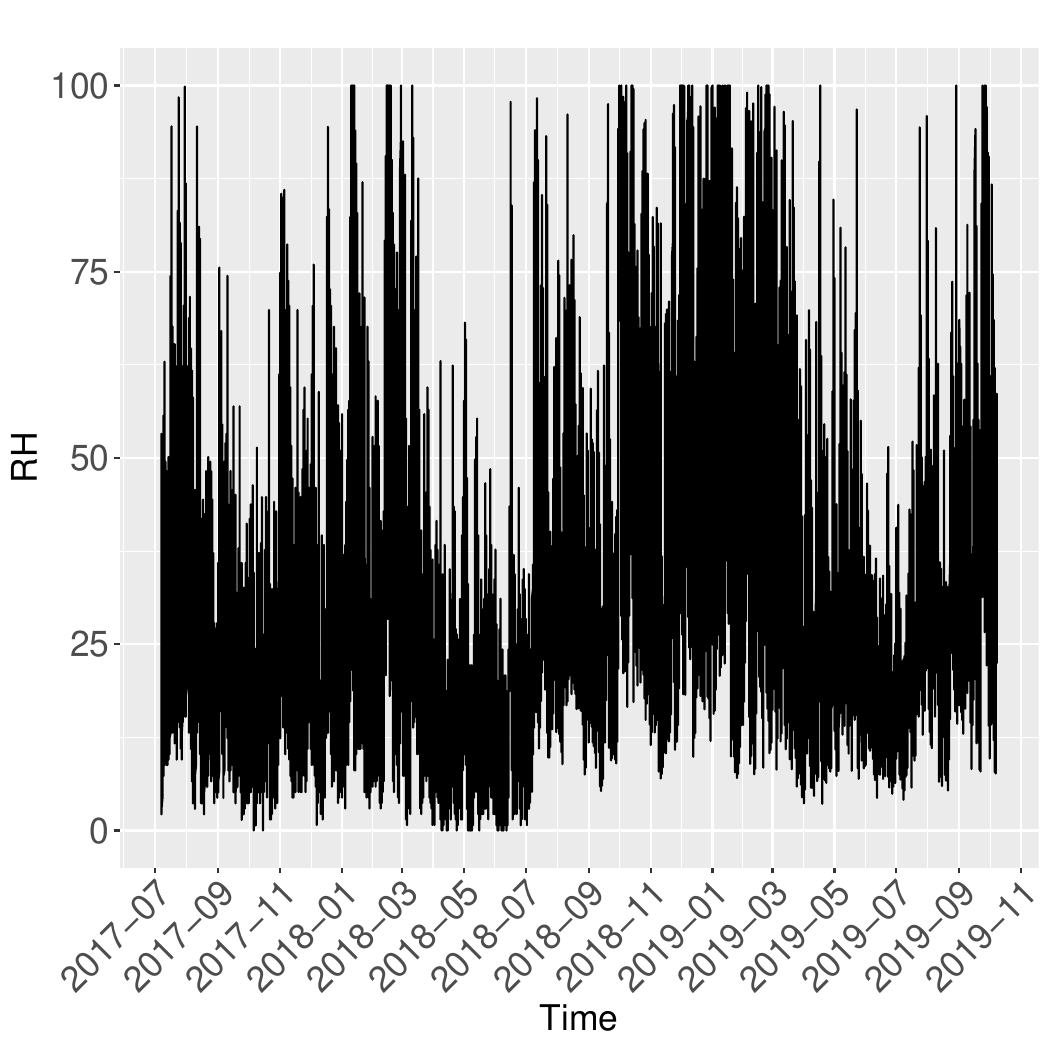}\\
(c) Temperature & (d) RH
\end{tabular}
\caption{Plot of outdoor time-varying covariates for the Arizona unit. Panels~(a) and~(b) present the profile and 3D perspective plots of solar irradiance, while panels~(c) and~(d) display the plots of temperature and relative humidity. The starting date of the covariate measurements is July 7, 2017. The unit of irradiance is W/$\text{m}^2$/nm, the unit of temperature is $\degreeC$, and the unit of RH is \%. }\label{fig:outdoor.cov.data.plot}
\end{center}
\end{figure}

Because of the high resolution and the inclusion of irradiance profiles, the outdoor environmental data are large in scale. Figure~\ref{fig:outdoor.cov.data.plot} illustrates the time-varying environmental variables for the Arizona unit. In panel (a), each curve represents $E(\lambda, \tau)$ for a given $\tau$, while panel (b) provides a 3D perspective plot of $E(\lambda, \tau)$, with the time axis showing the number of hours since the start of the experiments. The temperature exhibits a clear seasonal pattern, whereas RH shows substantial variability over time.

Note that for the outdoor data, only four samples were repeatedly measured for YI and chemical change. For the modeling and prediction tasks, all outdoor samples were included in the test set. In other words, the outdoor units were used to evaluate the prediction performance of both the statistical and DL models. The outdoor measurements also serve as an external validation for assessing whether the indoor acceleration mechanisms hold under real-world outdoor conditions. This design provides a stringent test of model generalizability, since the outdoor environment introduces greater variability and complexity than the controlled indoor settings. By comparing indoor-based predictions with outdoor observations, we can better understand the strengths and limitations of the proposed modeling framework.

%%%%%%%%%%%%%%%%%%%%%%%%%%%%%%%%%%%%%%%%%%%%%%%%%%%%%%%%%%%%%%%%%%%%%%%%%%%%%%%%%%%%%%%
\section{Statistical Approach for Building Predictive Models}\label{sec:statistical.approach}
%%%%%%%%%%%%%%%%%%%%%%%%%%%%%%%%%%%%%%%%%%%%%%%%%%%%%%%%%%%%%%%%%%%%%%%%%%%%%%%%%%%%%%%

In this section, we introduce statistical methods for developing predictive models of PPE degradation, while the data analysis is deferred to Section~\ref{sec:indoor.data.analysis}.

%%%%%%%%%%%%%%%%%%%%%%%%%%%%%%%%%%%%%%%%%%%%%%%%%%%%%%%%%%%%%%%%%%%%%%%%%%%%%%%%%%%%%%%
\subsection{General Path Model and Effective Dosage}
%%%%%%%%%%%%%%%%%%%%%%%%%%%%%%%%%%%%%%%%%%%%%%%%%%%%%%%%%%%%%%%%%%%%%%%%%%%%%%%%%%%%%%%

Let $y_{ij}$ denote the $j$th degradation measurement for unit $i$ (i.e., specimen $i$). The statistical model for the degradation measurements is
\begin{align}\label{eqn:gpm.main.model}
y_{ij} = D_i(t_{ij}) + \epsilon_{ij},
\end{align}
where $D_i(t_{ij})$ represents the general degradation path as a function of time $t_{ij}$ and experimental conditions. The deviation $\epsilon_{ij}$ captures variability not explained by $D_i(t_{ij})$. We assume $\epsilon_{ij}$ follows a normal distribution with mean zero and variance $\sigma_\epsilon^2$, that is, $\epsilon_{ij} \sim \NOR(0, \sigma_\epsilon^2)$.

The experimental conditions, including UV spectrum, intensity, temperature, and RH, are incorporated into the general path function $D_i(t)$. The UV radiation originates from the irradiance $E_i(\lambda)$ and passes through filter(s) $F_i(\lambda)$, resulting in the wavelength-specific intensity
$I_i(\lambda) = E_i(\lambda) \times F_i(\lambda)$.
The usual dosage from the lamp is computed as
\[
d_i(t) = \int_{0}^{t}\int_{\lambda} I_i(\lambda) d\lambda d\tau
       = t \int_{\lambda} I_i(\lambda) d\lambda.
\]
To account for the effects of both wavelength and intensity, we introduce the concept of effective dosage, defined as
\begin{align}\label{eqn:def.effective.dosage}
s_i(t) = \int_{0}^{t}\int_{\lambda} [I_i(\lambda)]^p  \phi(\lambda) d\lambda d\tau
       = t \times \int_{\lambda} [I_i(\lambda)]^p \phi(\lambda) d\lambda.
\end{align}

\begin{figure}
\begin{center}
\includegraphics[width=.85\textwidth]{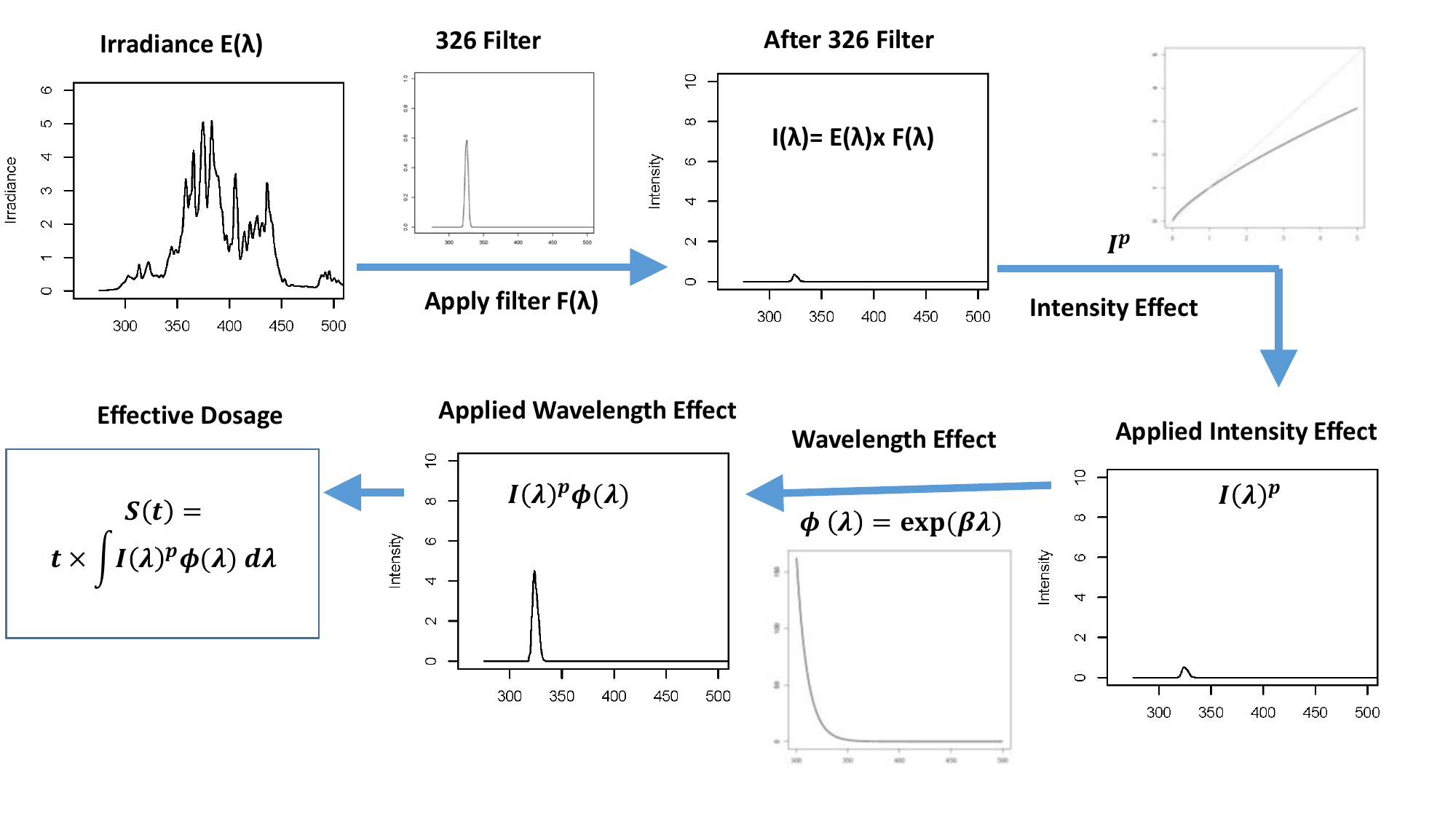}
\caption{Illustration of the procedure for computing the effective dose.}\label{fig:eff.dose.cmpt}
\end{center}
\end{figure}

The effect of wavelength is described by the function $\phi(\lambda)$, which allows different wavelengths to have different effects. Based on the literature (e.g., \citealt{Guetal2008}), a log-linear relationship is often used for the wavelength effect. Specifically, one can specify $\phi(\lambda) = \exp[\beta(\lambda - 354)]$. Here, 354$\nm$ is used as the baseline, meaning that the acceleration factor at 354$\nm$ is one. For the effect of intensity, a power-law relationship is often assumed. That is, the effect of intensity is specified as $[I(\lambda)]^p$, with power coefficient $p$. When $p=1$, the reciprocity law is satisfied.  Figure~\ref{fig:eff.dose.cmpt} illustrates the procedure for computing the effective dosage.

The general path is modeled as a function of the effective dosage,
$$D_i(t) = g[s_i(t)],$$
through a physically motivated parametric model $g(s)$. For polymeric materials, the log-logistic function is widely used as the general path model for $g(s)$ (e.g., \citealt{Duanetal2017}). According to \citet{Fockeetal2017}, this model arises as a solution to the Prout-Tompkins equation in kinetic modeling. Specifically,
$$
\frac{dg(s)}{ds} = \frac{A}{\sigma \exp(\nu)} \left[\frac{g(s)}{A}\right]^{1-\sigma} \left[1 - \frac{g(s)}{A}\right]^{1+\sigma},
$$
which has a solution
\begin{align}\label{eqn:log.logistic.form}
g(s) = \frac{A}{1+\exp\left[-\dfrac{\log(s)-\nu}{\sigma}\right]}.
\end{align}
The model in \eqref{eqn:log.logistic.form} offers flexibility to capture a wide range of degradation path shapes through the parameters $A$, $\nu$, and $\sigma$. Covariates can be incorporated into the model via the parameter $\nu$; see Section~\ref{sec:temp.rh.model} for further details.

With the functional form of $g(\cdot)$ specified in \eqref{eqn:log.logistic.form}, the model in \eqref{eqn:gpm.main.model} can be written as
\begin{align}\label{eqn:gpm.main.model.gfun}
y_{ij} = g[s_{i}(t_ij)] + \epsilon_{ij}=\frac{A}{1+\exp\left\{-\dfrac{\log[s_i(t_{ij})]-\nu}{\sigma}\right\}} + \epsilon_{ij},
\end{align}
for an increasing path. The parameters in model \eqref{eqn:gpm.main.model.gfun} have meaningful interpretations. The parameter $A$ represents the ultimate degradation. Define $\eta = \exp(\nu)$ and $\gamma = 1/\sigma$. Here, $\eta$ gives the half-degradation effective dosage, i.e., the amount of effective dosage required for the degradation to reach $0.5A$. The parameter $\gamma$ characterizes the steepness of the damage curve. For example, the slope at $s(t)=\eta$ is $A\gamma/(4\eta)$. Thus, a larger $\gamma$ corresponds to a steeper curve.

%%%%%%%%%%%%%%%%%%%%%%%%%%%%%%%%%%%%%%%%%%%%%%%%%%%%%%%%%%%%%%%%%%%%%%%%%%%%%%%%%%%%%%%%%%%%%%%%%%%
\subsection{Modeling Temperature and Humidity Effects}\label{sec:temp.rh.model}
%%%%%%%%%%%%%%%%%%%%%%%%%%%%%%%%%%%%%%%%%%%%%%%%%%%%%%%%%%%%%%%%%%%%%%%%%%%%%%%%%%%%%%%%%%%%%%%%%%%
To model the effects of temperature and RH, we incorporate them into $\nu$ through the function $\nu(\xvec_i)$, where $\xvec_i$ contains the information on $\TempC_i$ and $\RH_i$. That is, in \eqref{eqn:gpm.main.model.gfun}, $\nu$ is replaced by $\nu(\xvec_i)$. In the reliability literature, the Arrhenius relationship is commonly used to describe the temperature acceleration factor. That is,
\begin{align*}
f(\TempC_i) = \exp\left[\beta_{\temp} \times \left(\frac{11605}{\TempC_i + 273.15}-\frac{11605}{85 + 273.15}\right)\right].
\end{align*}
Here, the temperature is expressed in Kelvin and we use 85$\degreeC$ as the baseline. In the PV literature, there are few functional forms available for modeling RH effects. A power-law relationship was used in \citet{Kaayaetal2021}. Our data include only two RH levels (0$\pect$ and 60$\pect$). Therefore, a relatively simple form is selected for the RH effect. In this paper, it is modeled as $f(\RH_i) = (1 + \RH_i)^{\beta_{\rh}}$. Since $\RH_i$ is bounded between 0 and 1, we add 1 to ensure that 0$\pect$ RH serves as the baseline.

Let $\xtemp_i = \left[11605/(\TempC_i + 273.15) - 11605/(85 + 273.15)\right]$ and $\xrh_i = \log(1+\RH_i)$ be the transformed variables for temperature and RH, respectively. Based on the above discussion, we model $\nu(\xvec_i)$ as
\begin{align}\label{eqn:nu.xvec}
\nu(\xvec_i) = \mu + \beta_{\text{temp}}\cdot \xtemp_i + \beta_{\text{rh}}\cdot \xrh_i.
\end{align}
In summary, the model for an increasing degradation path is
\begin{align}\label{eqn:gpm.main.model.gfun.covariates}
D_i(t) =D_i(t;\thetavec)= g[s_i(t)] = \frac{A}{1+\exp\left\{-\dfrac{\log[s_i(t)]-\nu(\xvec_i)}{\sigma}\right\}}.
\end{align}
The model parameters are
$\thetavec = (A, \mu, \sigma, \beta, p, \beta_{\text{temp}}, \beta_{\text{rh}}, \sigma_{\epsilon}^2)\tran.$

%%%%%%%%%%%%%%%%%%%%%%%%%%%%%%%%%%%%%%%%%%%%%%%%%%%%%%%%%%%%%%%%%%%%%%%%%%%%%%%%%%%%%%%%%%%%%%%%%%%
\subsection{Parameter Estimation and Degradation Prediction}
%%%%%%%%%%%%%%%%%%%%%%%%%%%%%%%%%%%%%%%%%%%%%%%%%%%%%%%%%%%%%%%%%%%%%%%%%%%%%%%%%%%%%%%%%%%%%%%%%%%

We use maximum likelihood estimation. The log-likelihood function is
\begin{align}\label{eqn:gpm.loglik}
\loglik(\thetavec) = -\frac{1}{2}\sum_{i}n_i\log(2\pi) - \sum_{i}\frac{n_i}{2}\log(\sigma_{\epsilon}^2)-\sum_{i,j}\frac{[y_{ij} - D_i(t_{i,j}; \thetavec)]^2}{2\sigma_{\epsilon}^2},
\end{align}
where $n_i$ is the number of observations from unit $i$. It is clear that there is no closed-form solution for the maximum likelihood estimate (MLE) of $\thetavec$, and numerical algorithms are therefore required for the optimization. Due to the complexity of the log-likelihood function, we employ the Nelder-Mead algorithm (\citealt{nelder1965simplex}), which is a function-value-based method. Although slower than gradient-based approaches, it is generally more robust. In contrast, the MLE of the variance $\sigma_{\epsilon}^2$ has a closed-form solution, which is given by
$$
\widehat{\sigma}_{\epsilon}^2 = \frac{\sum_{i,j}[y_{ij} - D_i(t_{i,j}; \thetavechat)]^2}{\sum_i n_i}.
$$

The MLE is denoted as $\thetavechat = (\Ahat, \muhat, \sigmahat, \betahat, \phat, \betahat_{\temp}, \betahat_{\rh}, \sigmahat_{\epsilon}^2)\tran.$ The inference is based on asymptotic theory, that is,
$$\thetavechat \approxsim \NOR(\thetavechat, \Sigma_{\thetavechat}).$$
Here, $\Sigma_{\thetavechat} = I_{\thetavechat}^{-1}$ denotes the inverse of the observed information matrix evaluated at $\thetavechat$.

To predict the mean degradation for a new indoor unit under a new experimental condition, we substitute the MLEs into the model. The predicted effective dosage is given by
\begin{align}\label{eqn:def.effective.dosage.estimated}
s_{\new}(t;\thetavechat) = \int_{0}^{t}\int_{\lambda} [I_{\new}(\lambda)]^{\phat},\widehat{\phi}(\lambda), d\lambda, d\tau,
\end{align}
where $\widehat{\phi}(\lambda)= \exp[\widehat{\beta}(\lambda - 354)]$. The corresponding predicted degradation path is
\begin{align}\label{eqn:indoor.predicted.path}
D_{\new}(t;\thetavechat) = g[s_{\new}(t; \thetavechat); \xvec_{\new}, \thetavechat]
=\frac{\Ahat}{1+\exp\left\{-\dfrac{\log[s_{\new}(t;\thetavechat)]
-\nu(\xvec_{\new};\thetavechat)}{\sigmahat}\right\}},
\end{align}
where $s_{\new}(t;\thetavechat)$ is defined in \eqref{eqn:def.effective.dosage.estimated}, and $$\nu(\xvec_{\new}; \thetavechat) = \muhat + \betahat_{\temp}\xtemp_{\new} + \betahat_{\rh}\xrh_{\new}.$$

We use a simulation-based approach to compute the confidence interval. Although the delta method can be applied to derive the large-sample variance for a function of $\thetavechat$, the simulation-based approach is equivalent and simpler for practitioners to implement. With modern computing power, it can also be carried out efficiently. We present the procedure in Algorithm~\ref{algo:pci.indoor.pred}, which computes the pointwise confidence intervals for the predicted path of a new indoor unit.

\begin{algorithm}{$100(1-\alpha)\pect$ Pointwise Confidence Intervals for Indoor Prediction}\label{algo:pci.indoor.pred}
\begin{enumerate}[label=(\arabic*)]
\item Draw a simulated parameter vector $\thetavechat^{\ast}$ from $\NOR(\thetavechat, \Sigma_{\thetavechat})$.
\item Using $\thetavechat^{\ast}$, compute the predicted degradation path $D_{\new}(t;\thetavechat^{\ast})$ as defined in \eqref{eqn:indoor.predicted.path}.
\item Repeat Steps 1 and 2 for $b=1,\dots,B$ to obtain $\{D_{\new}(t;\thetavechat^{\ast b})\}_{b=1}^B$.
\item Construct the confidence interval for $D_{\new}(t)$ by taking the $\alpha/2$ and $1-\alpha/2$ quantiles of the simulated values $\{D_{\new}(t;\thetavechat^{\ast b})\}_{b=1}^{B}$.
\end{enumerate}
\end{algorithm}

Confidence intervals for other functions of $\thetavec$, such as $\phi(\lambda)$, can be obtained in the same manner as described in Algorithm~\ref{algo:pci.indoor.pred}. This approach eliminates the need for mathematical derivations required by the delta method and is more user-friendly for practitioners in material science. Regarding the number of repeats $B$, we recommend using $B = 5{,}000$ and repeating the procedure to check the stability of the intervals. One can increase it to $10{,}000$ if needed.

%%%%%%%%%%%%%%%%%%%%%%%%%%%%%%%%%%%%%%%%%%%%%%%%%%%%%%%%%%%%%%%%%%%%%%%%%%%%%%%%%%%%%%%
\section{Indoor Data Analysis}\label{sec:indoor.data.analysis}
%%%%%%%%%%%%%%%%%%%%%%%%%%%%%%%%%%%%%%%%%%%%%%%%%%%%%%%%%%%%%%%%%%%%%%%%%%%%%%%%%%%%%%%
In this section, we present results on the indoor data analysis.

%%%%%%%%%%%%%%%%%%%%%%%%%%%%%%%%%%%%%%%%%%%%%%%%%%%%%%%%%%%%%%%%%%%%%%%%%%%%%%%%%%%%%%%
\subsection{Analysis of Yellowness Index}
%%%%%%%%%%%%%%%%%%%%%%%%%%%%%%%%%%%%%%%%%%%%%%%%%%%%%%%%%%%%%%%%%%%%%%%%%%%%%%%%%%%%%%%

\begin{figure}[h]
\begin{center}
\begin{tabular}{cc}
\includegraphics[width=.35\textwidth]{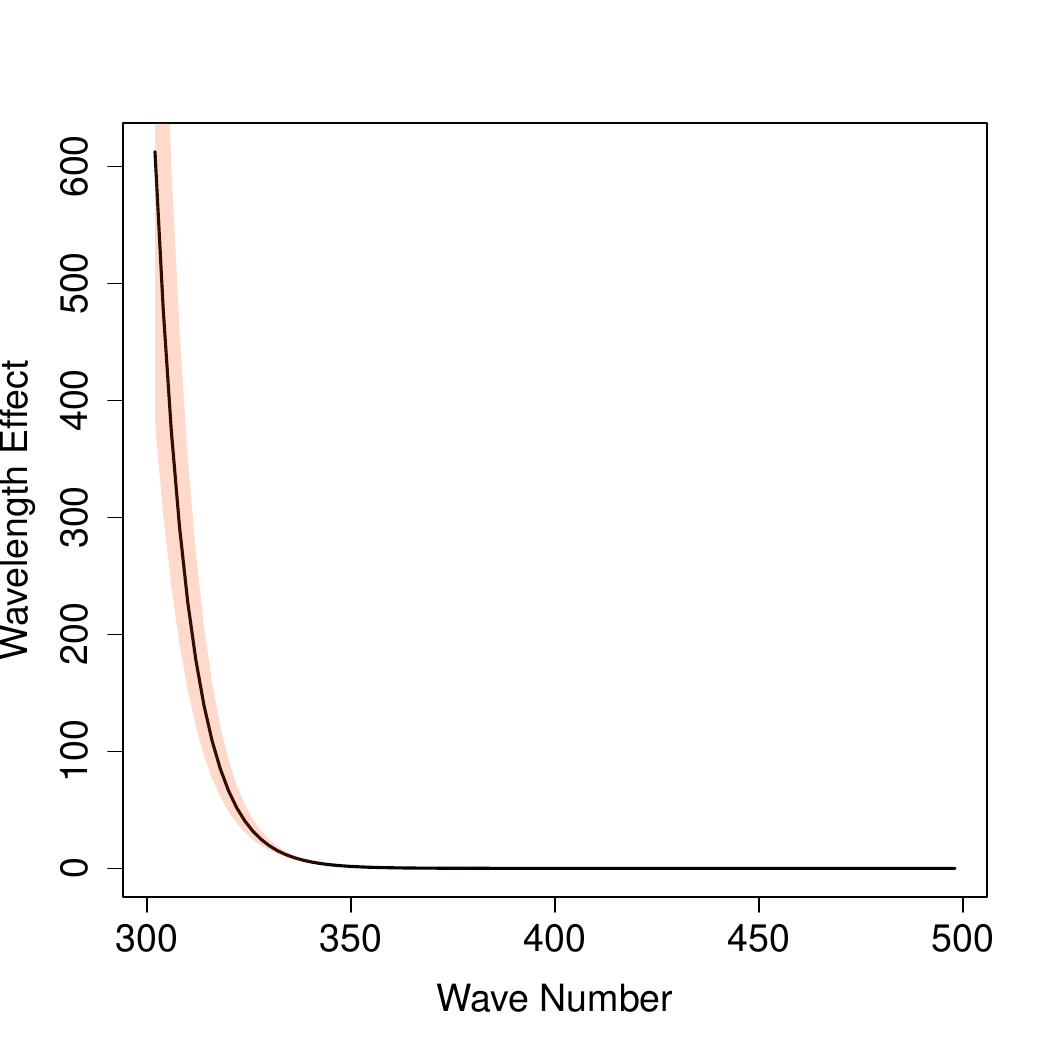}&
\includegraphics[width=.35\textwidth]{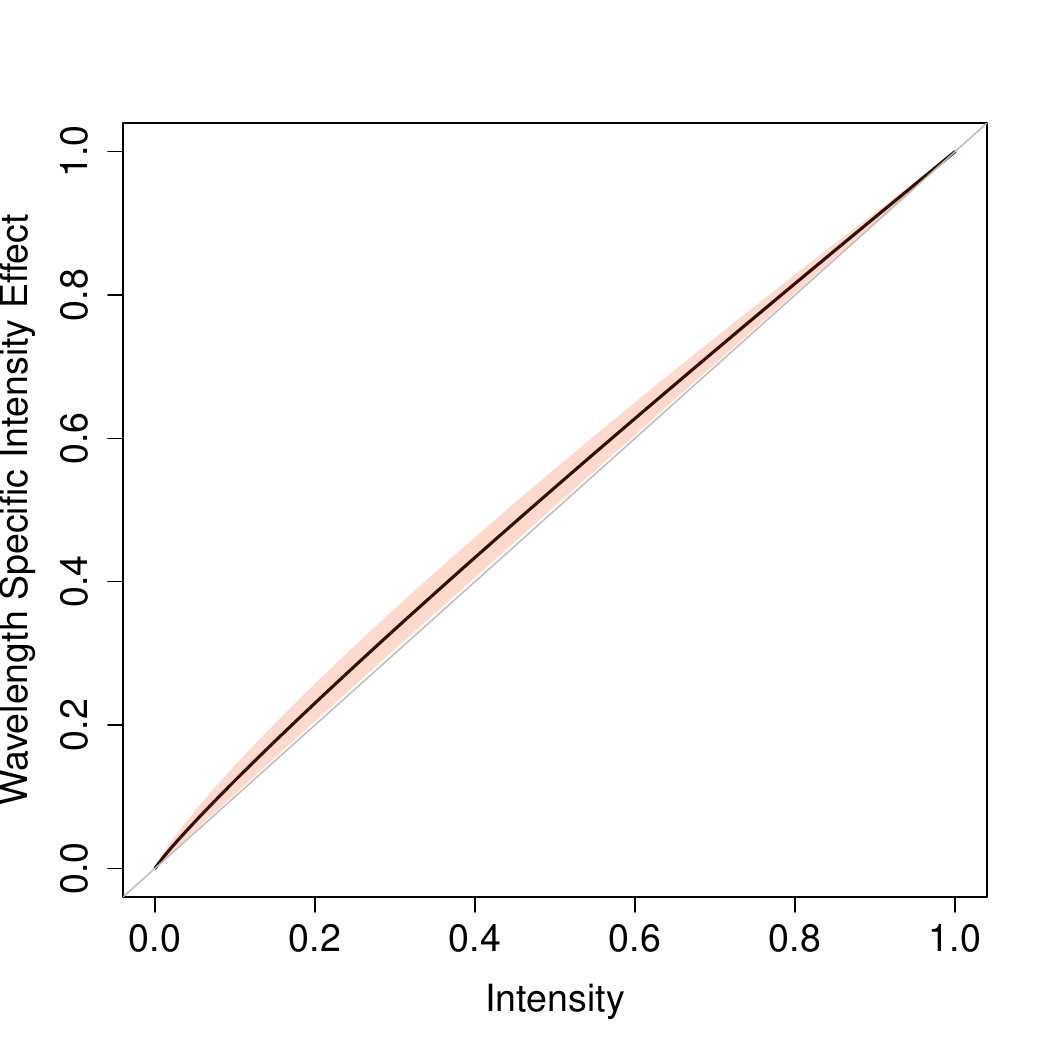}\\[-0.75em]
(a) Wavelength & (b) Intensity \\[-0.75em]
\includegraphics[width=.35\textwidth]{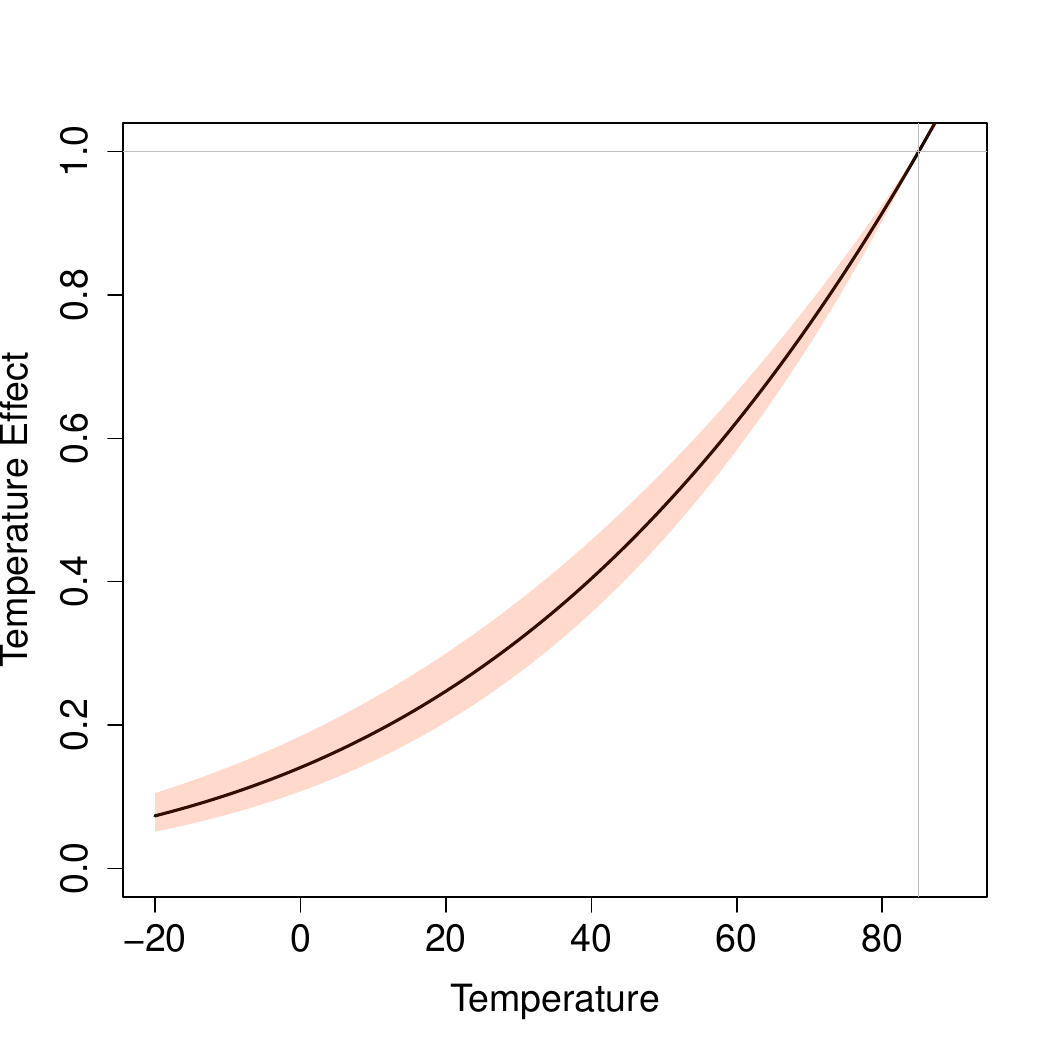}&
\includegraphics[width=.35\textwidth]{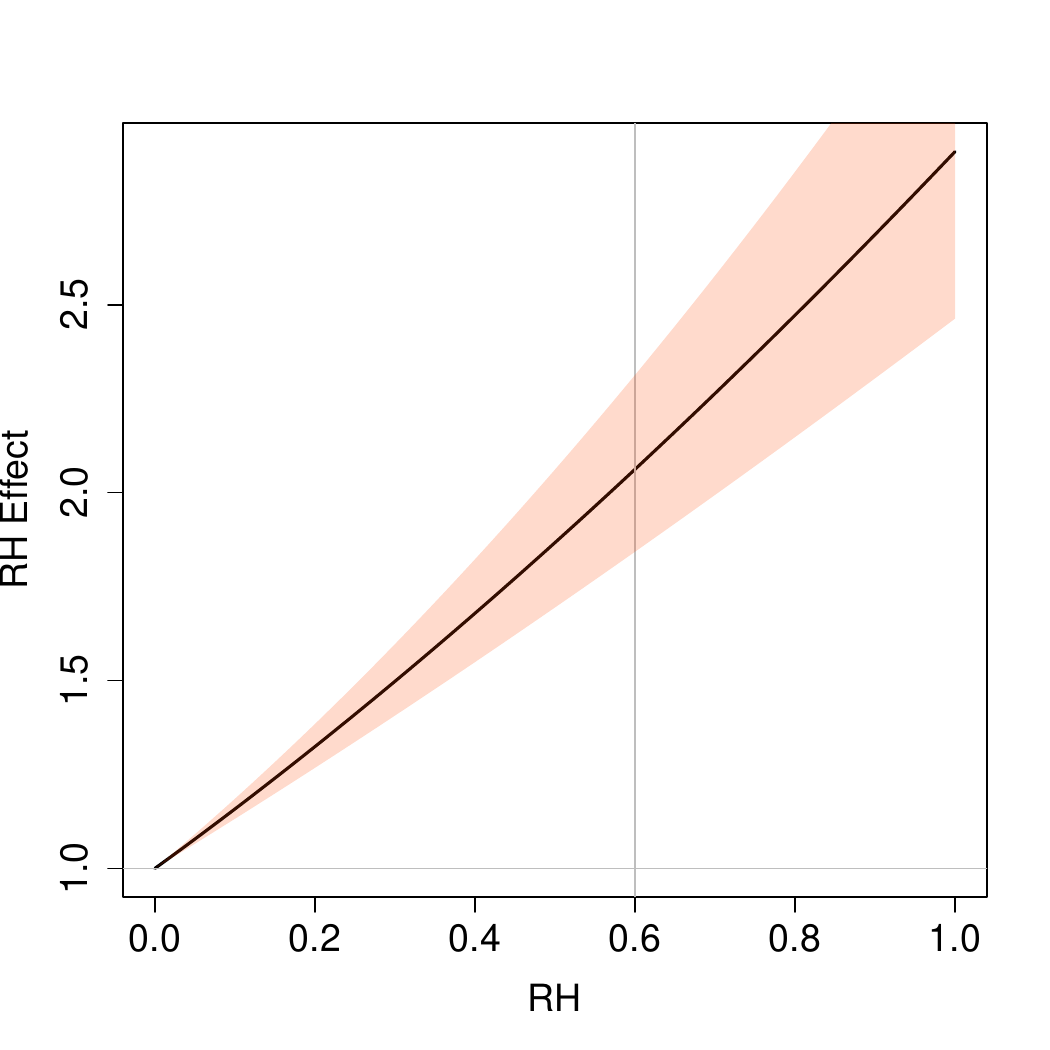}\\[-0.75em]
(c) Temperature  & (d) RH
\end{tabular}
\caption{Plots of the effect functions for the environmental variables based on the indoor degradation data on yellowness index. The black lines represent the estimates, and the shaded areas indicate the 95$\pect$ pointwise confidence bands. }\label{fig:eff.fun.plot.yi}
\end{center}
\end{figure}

The YI is increasing during exposure, and thus the model in~\eqref{eqn:gpm.main.model.gfun} can be applied directly. The training set specified in Table~\ref{tab:experiment.setup.YI} is used to obtain the MLE. Table~\ref{tab:YI.CR.par.est} reports the parameter estimates for the degradation models fitted to the yellowness index data. Figure~\ref{fig:eff.fun.plot.yi} presents the estimated effect functions, with shaded regions indicating pointwise confidence bands that quantify statistical uncertainty. In particular, Figure~\ref{fig:eff.fun.plot.yi}(a) shows a strong wavelength effect, with damage increasing exponentially at shorter UV wavelengths (around 306$\nm$). For the intensity effect, the estimate $p$ is close to 1, indicating that the reciprocity law holds. For the temperature effect, the activation energy is $0.195\,\text{eV}\times96.485=18.814\,\text{kJ/mol}$. The estimates also suggest a moderate RH effect, although the uncertainty is high because only two RH levels are available in the data.

\begin{figure}
\begin{center}
\begin{tabular}{c}
\includegraphics[width=.95\textwidth]{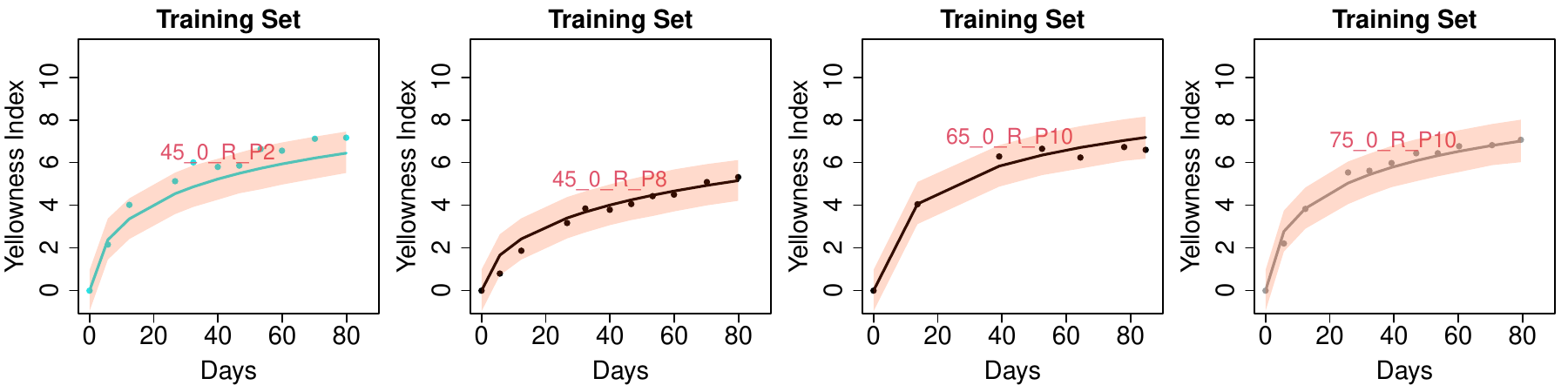}\\
\includegraphics[width=.95\textwidth]{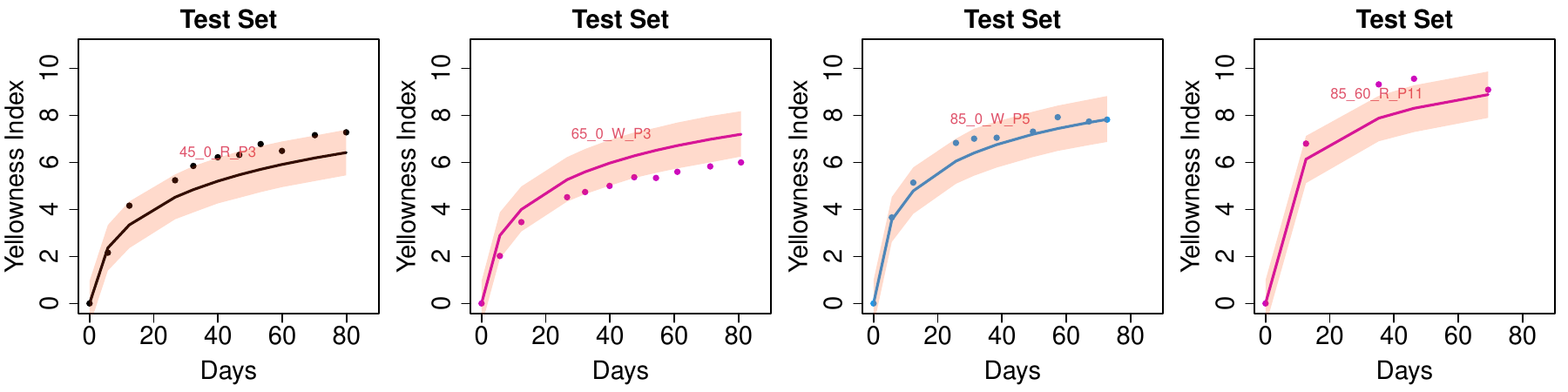}
\end{tabular}
\caption{Plot of the fitted (predicted) degradation paths for the yellowness index of eight representative units. The solid line represents the fitted (predicted) path, while the dots represent the observed data. The shaded area indicates the 90$\pect$ pointwise confidence interval. }\label{fig:fitted.plot.YI}
\end{center}
\end{figure}

Because the main focus of this paper is on degradation prediction, we compute the predicted degradation for the test set specified in Table~\ref{tab:experiment.setup.YI}. Figure~\ref{fig:fitted.plot.YI} displays the fitted (predicted) degradation paths for the yellowness index of eight representative units. The top row shows the fitted paths for units in the training set, and the bottom row shows those for the test set. The solid lines represent the fitted (predicted) paths, while the dots correspond to the observed data. The shaded regions indicate the pointwise confidence intervals, computed using Algorithm~\ref{algo:pci.indoor.pred}. Overall, the parametric model yields predictions with good accuracy.

\begin{table}
\caption{Parameter estimates for the degradation models fitted to degradation data on yellowness index and chemical change.}\label{tab:YI.CR.par.est}
\begin{center}
\begin{tabular}{c|c|cr|rc}\hline\hline
\multirow{2}{*}{Parameter} & \multirow{2}{*}{Interpretation }&&
\multicolumn{2}{c}{Estimate} & \\\cline{4-5}
&          & &    Yellowness Index & Chemical Change&     \\\hline
$A$       & ultimate (initial) degradation  &   &   12.048    &      1.010   &  \\\hline
$\eta=\exp(\mu)$    & half-degradation dosage & &   27.938    &   1916.010   &  \\\hline
$\gamma=1/\sigma$  & steepness              &   &    0.583    &      0.381   &  \\\hline
$\beta$   & wavelength effect               &   & $-$0.123    &   $-$0.180   &  \\\hline
$p$       & intensity effect                &   &    0.912    &      1.289   &  \\\hline
$\beta_{\temp}$ & temperature effect          & &    0.195    &      0.155   &  \\\hline
$\beta_{\rh}$ & RH effect                     & & $-$1.540    &   $-$4.794   &  \\\hline
$\sigma_{\epsilon}^2$ & error variance      &   &   0.3458    &     0.0028   &  \\\hline
\hline
\end{tabular}
\end{center}
\end{table}

%%%%%%%%%%%%%%%%%%%%%%%%%%%%%%%%%%%%%%%%%%%%%%%%%%%%%%%%%%%%%%%%%%%%%%%%%%%%%%%%%%%%%%%
\subsection{Analysis of Chemical Change}

For the chemical change (i.e., the peak ratio of 1245 vs. 1410), the degradation path is decreasing. To accommodate this behavior, we modify the model in~\eqref{eqn:gpm.main.model.gfun} as follows:
$$
D(t)=\frac{A}{1+\exp\left\{\dfrac{\log[s_i(t_{ij})]-\nu(\xvec_i)}{\sigma}\right\}},
$$
where $\nu(\xvec_i)$ is defined in \eqref{eqn:nu.xvec}.  The training set specified in Table~\ref{tab:CR.sample.list} is used to obtain the MLE.

\begin{figure}[h]
\begin{center}
\begin{tabular}{cc}
\includegraphics[width=.35\textwidth]{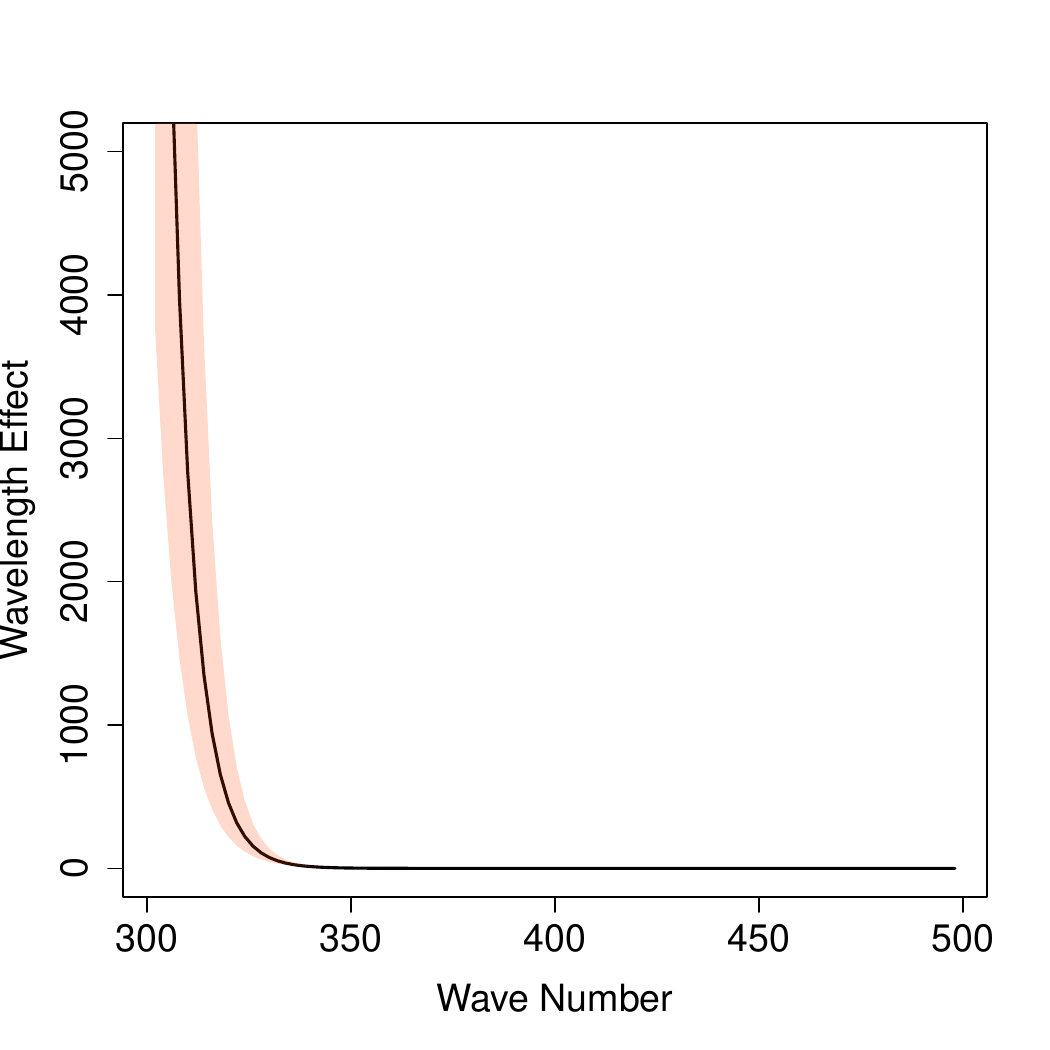}&
\includegraphics[width=.35\textwidth]{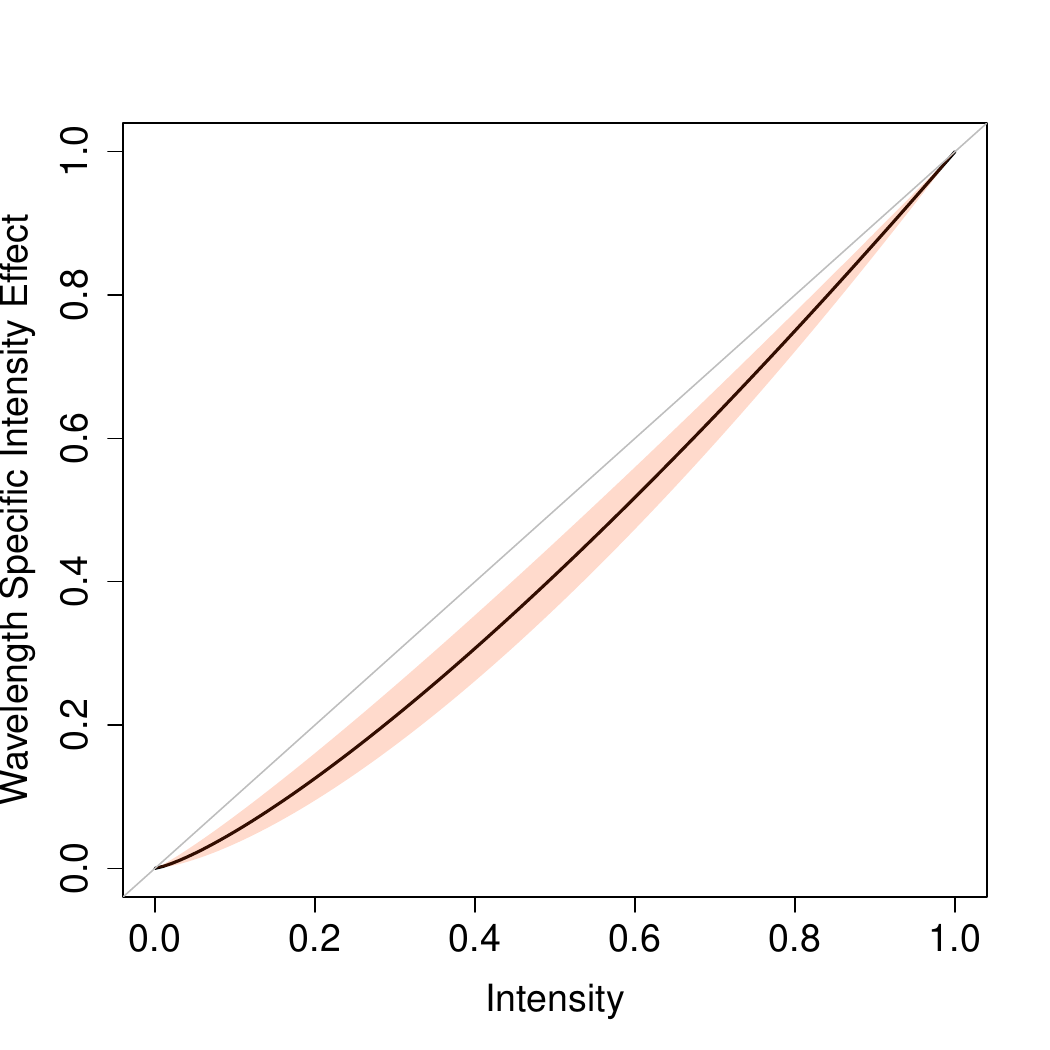}\\[-0.75em]
(a) Wavelength  & (b) Intensity\\[-0.75em]
\includegraphics[width=.35\textwidth]{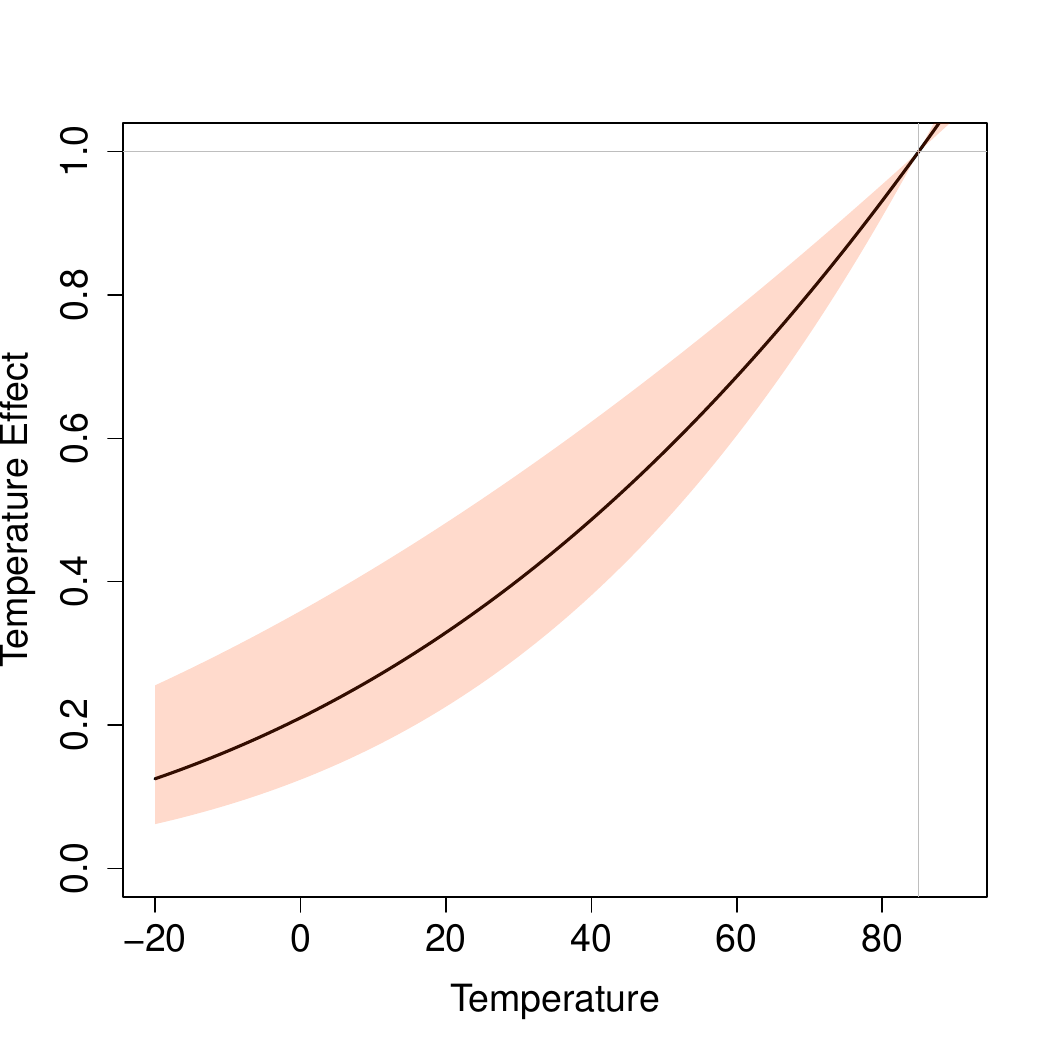}&
\includegraphics[width=.35\textwidth]{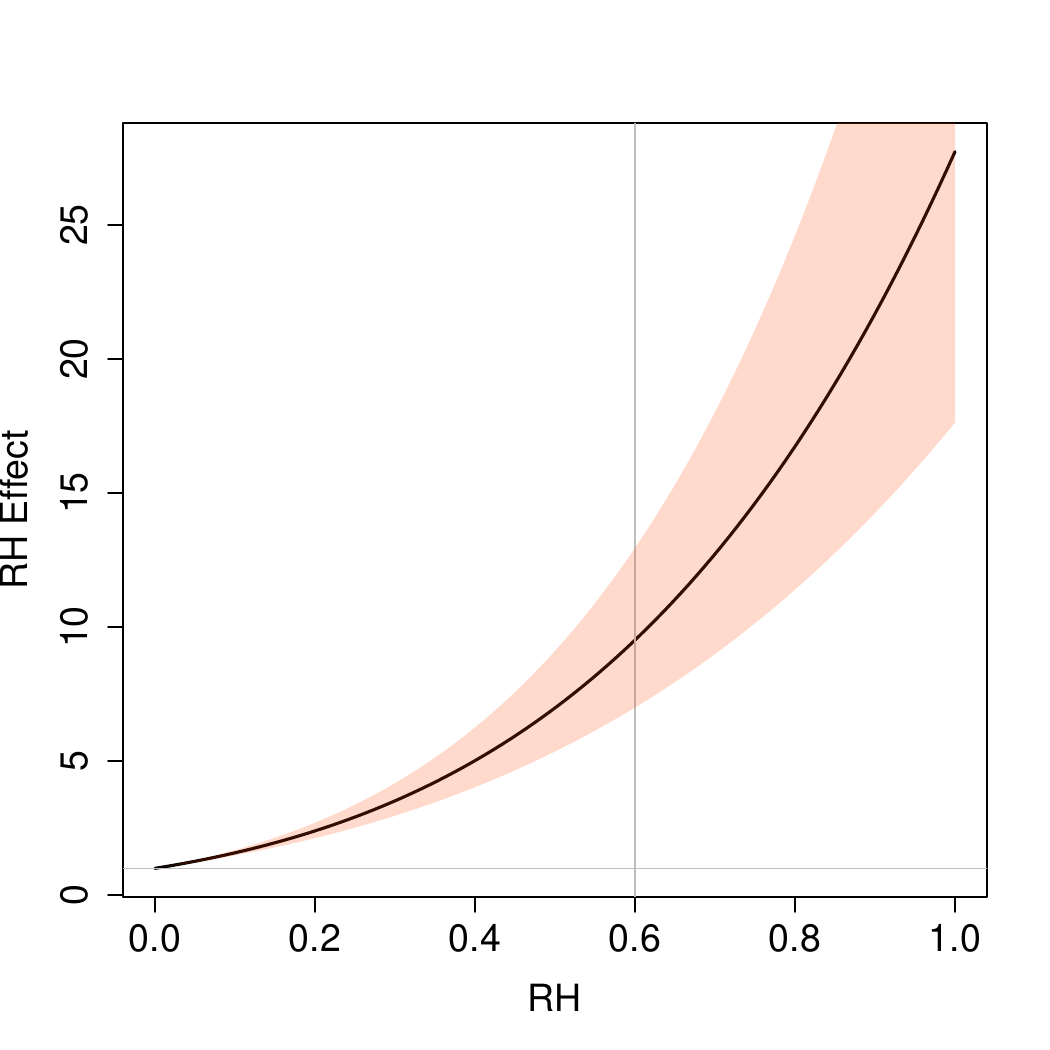}\\[-0.75em]
(c) Temperature  & (d) RH
\end{tabular}
\caption{Plots of the effect functions for the environmental variables based on the indoor degradation data on chemical change. The black lines represent the estimates, and the shaded areas indicate the 95$\pect$ pointwise confidence bands.}\label{fig:eff.fun.plot.chemical.change}
\end{center}
\end{figure}

Table~\ref{tab:YI.CR.par.est} reports the parameter estimates for the degradation models fitted to the chemical change data. Figure~\ref{fig:eff.fun.plot.chemical.change} shows the estimated effect functions, with shaded regions representing pointwise confidence bands. Similar to the yellowness index, wavelength exhibits a strong effect on degradation. The estimated $p$ is greater than 1, and the confidence band in Figure~\ref{fig:eff.fun.plot.chemical.change}(b) indicates it is significantly different from 1, suggesting that the reciprocity law may not hold for chemical change. For temperature, the activation energy is $0.155\,\text{eV}\times96.485=14.955\, \text{kJ/mol}$. The estimates also indicate a strong RH effect, though the large uncertainty highlights the challenge of estimating this effect.

\begin{figure}
\begin{center}
\begin{tabular}{c}
\includegraphics[width=.95\textwidth]{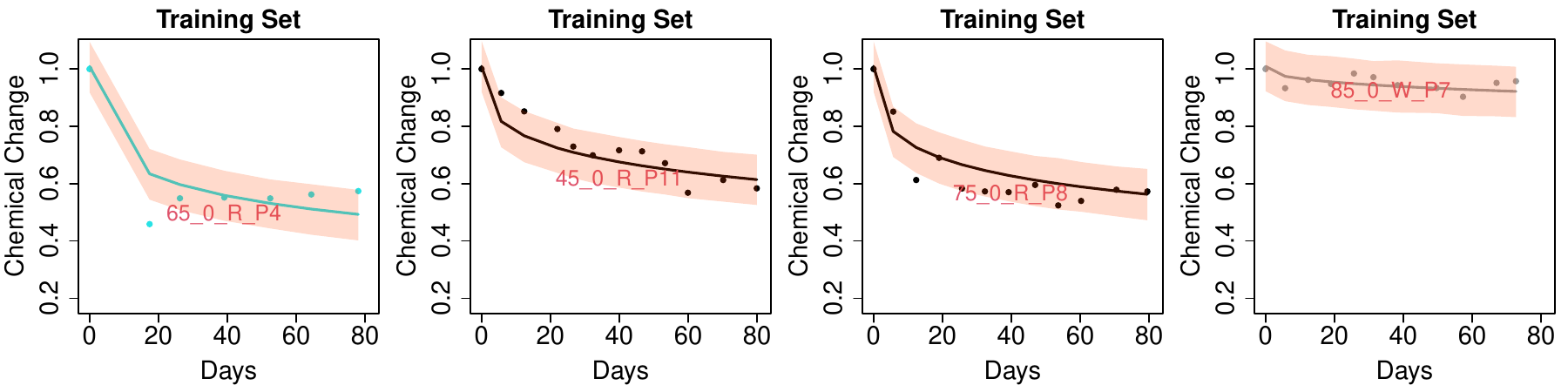}\\
\includegraphics[width=.95\textwidth]{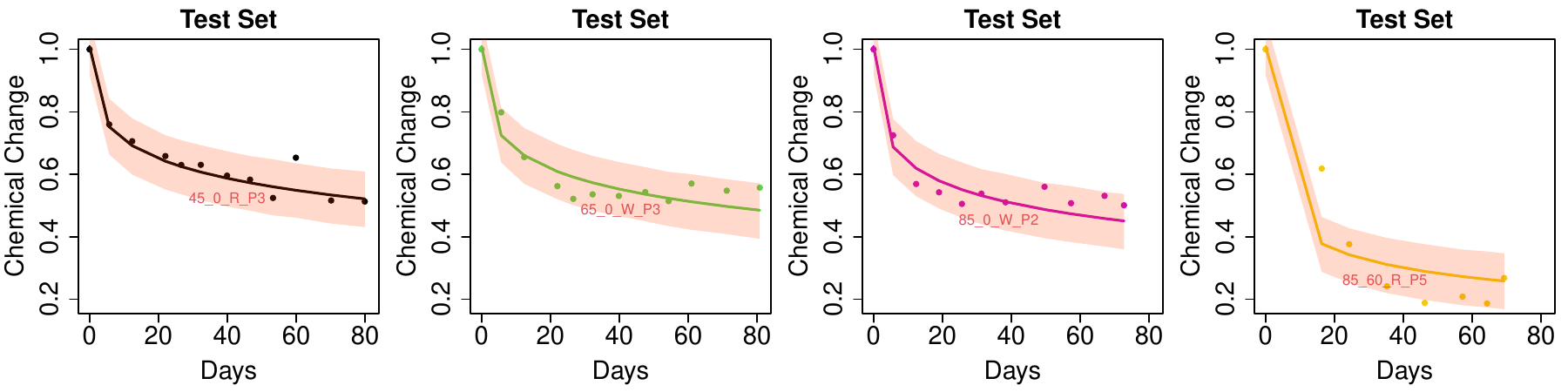}
\end{tabular}
\caption{Plot of the fitted (predicted) degradation paths for the chemical change of eight representative units. The solid line represents the fitted (predicted) path, while the dots represent the observed data. The shaded area indicates the 90$\pect$ pointwise confidence bands.}\label{fig:fitted.plot.chemical.change}
\end{center}
\end{figure}

For prediction, we compute the predicted paths for the test set specified in Table~\ref{tab:CR.sample.list}. Figure~\ref{fig:fitted.plot.chemical.change} illustrates the fitted (predicted) degradation paths for the chemical change of eight representative units. The top row shows the fitted paths for units in the training set, while the bottom row shows those for the test set. Similar to the results for the yellowness index, both the fitted and predicted degradation paths demonstrate good accuracy. In summary, the statistical model involves estimating eight parameters, and the 55 samples with repeated measurements over time provide enough information for this purpose. The results from both the training and test sets indicate that the model performs reasonably well.

%%%%%%%%%%%%%%%%%%%%%%%%%%%%%%%%%%%%%%%%%%%%%%%%%%%%%%%%%%%%%%%%%%%%%%%%%%%%%%%%%%%%%%%
\section{Statistical Approach for Outdoor Predictions}\label{sec:outdoor.prediction}
%%%%%%%%%%%%%%%%%%%%%%%%%%%%%%%%%%%%%%%%%%%%%%%%%%%%%%%%%%%%%%%%%%%%%%%%%%%%%%%%%%%%%%%
\subsection{Outdoor Prediction and Uncertainty Quantification}
%%%%%%%%%%%%%%%%%%%%%%%%%%%%%%%%%%%%%%%%%%%%%%%%%%%%%%%%%%%%%%%%%%%%%%%%%%%%%%%%%%%%%%%%%%%%%%%%%%%
For outdoor prediction, it is necessary to account for time-varying covariates. This is achieved through the use of the effective dosage formulation. To incorporate time-varying environmental variables into outdoor degradation modeling, the effective dosage is extended as
\begin{align}\label{eqn:outdoor.effective.dosage.cmpt}
s(t;\thetavec)=\int_{0}^{t}\exp\left[\beta_{\temp}\cdot\frac{11605}{\TempK(\tau)}\right][1+\RH(\tau)]^{\beta_{\rh}}\int_{\lambda}[E(\lambda, \tau)]^p\phi(\lambda),d\lambda,d\tau.
\end{align}

For the outdoor prediction model, the inputs to the effective dosage formulation consist of outdoor irradiance, temperature, and RH data as functions of time, together with the model parameters estimated from the indoor training set. The effective dosage $s(t;\thetavec)$ is then computed as a function of time. Using $s(t;\thetavec)$, the outdoor degradation is expressed as
\begin{align}\label{eqn:outdoor.deg.cmpt}
D(t; \thetavec)=\frac{A}{1+\exp\left\{-\dfrac{\log[s(t)]-\mu}{\sigma}\right\}},
\end{align}
where $s(t)=s(t;\thetavec)$ is defined in \eqref{eqn:outdoor.effective.dosage.cmpt}. With some mathematical derivation, it can be shown that when both temperature and RH are time-invariant, the model in \eqref{eqn:outdoor.deg.cmpt} reduces to the model in \eqref{eqn:gpm.main.model.gfun}, which handles time-invariant covariates.

For outdoor prediction, the results are conditional on the observed environmental factors (i.e., UV exposure, temperature, and RH). Point predictions can be obtained by plugging the MLEs and time-varying covariates into the model. Uncertainty quantification is more challenging due to the complexity of the outdoor data structure; therefore, we rely on a simulation-based approach, which is straightforward to implement and is detailed in Algorithm~\ref{algo:pci.outdoor.pred}. This framework provides a practical way to generate prediction with uncertainty quantification.

\begin{algorithm}{$100(1-\alpha)\pect$ Pointwise Confidence Intervals for Outdoor Prediction}\label{algo:pci.outdoor.pred}
\begin{enumerate}[label=(\arabic*)]
\item Draw a simulated parameter vector $\thetavechat^{\ast}$ from $\NOR(\thetavechat, \Sigma_{\thetavechat})$.

\item Using $\thetavechat^{\ast}$ and covariate history, compute the effective dosage $s(t;\thetavechat^{\ast})$ using \eqref{eqn:outdoor.effective.dosage.cmpt}.

\item Compute the predicted degradation path $D(t; \thetavechat^{\ast})$ as in \eqref{eqn:outdoor.deg.cmpt}.

\item Repeat Steps 1-3 for $b=1,\dots,B$ to obtain $\{D(t;\thetavechat^{\ast b})\}_{b=1}^B$.

\item Construct the confidence interval for $D(t)$ by taking the $\alpha/2$ and $1-\alpha/2$ quantiles of the simulated values $\{D(t;\thetavechat^{\ast b})\}_{b=1}^{B}$.
\end{enumerate}
\end{algorithm}
%%%%%%%%%%%%%%%%%%%%%%%%%%%%%%%%%%%%%%%%%%%%%%%%%%%%%%%%%%%%%%%%%%%%%%%%%%%%%%%%%%%%%%%%%%%%%%%%%%%
\subsection{Outdoor Prediction Results}
%%%%%%%%%%%%%%%%%%%%%%%%%%%%%%%%%%%%%%%%%%%%%%%%%%%%%%%%%%%%%%%%%%%%%%%%%%%%%%%%%%%%%%%%%%%%%%%%%%%
\begin{figure}
\begin{center}
\begin{tabular}{c}
\includegraphics[width=.95\textwidth]{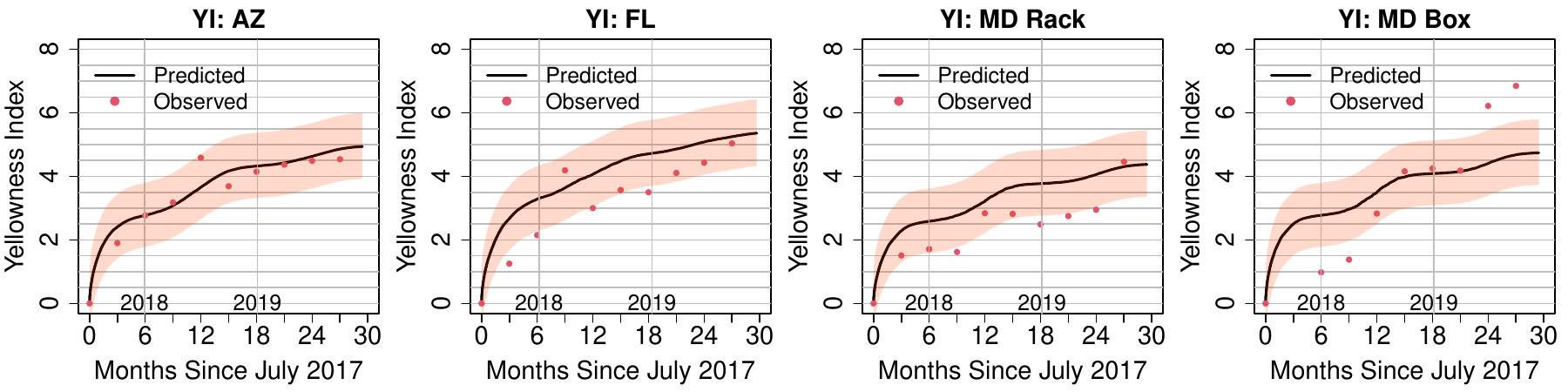}\\
\includegraphics[width=.95\textwidth]{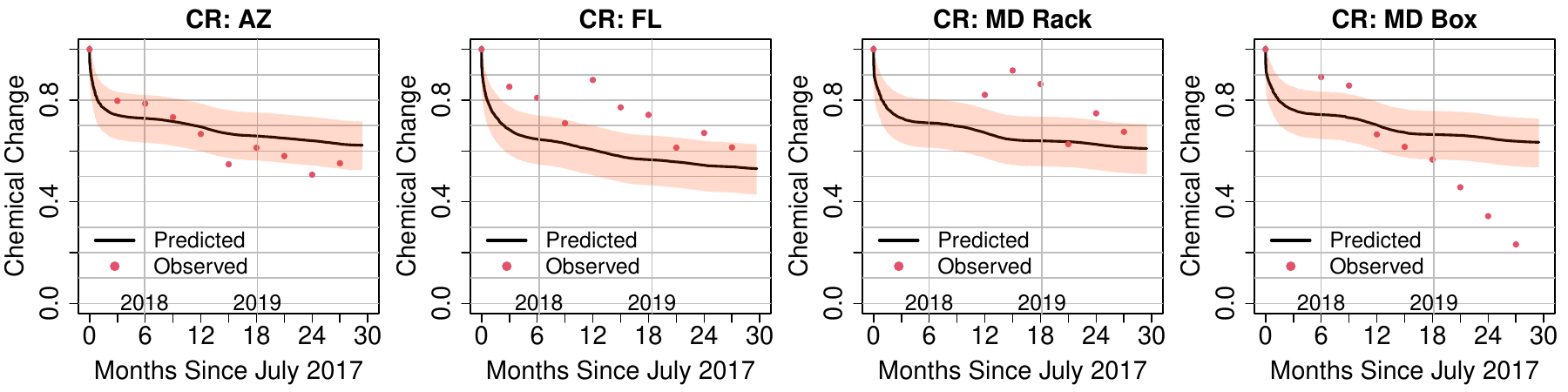}
\end{tabular}
\caption{Plot of the predicted degradation paths for the yellowness index and chemical change of the four outdoor units as functions of time. The shaded area represents the 90$\pect$ pointwise confidence intervals.}\label{fig:out.pred.res}
\end{center}
\end{figure}

Figure~\ref{fig:out.pred.res} presents the predicted degradation paths for the yellowness index and chemical change of the four outdoor units as functions of time, which is based on the model in \eqref{eqn:outdoor.deg.cmpt}. The shaded regions represent the pointwise confidence intervals. Note that the outdoor data were not used in training the model. Therefore, the outdoor predictions provide a validation of the accelerated degradation experiments conducted indoors. From Figure~\ref{fig:out.pred.res}, the yellowness index predictions align well with the observations for the AZ, FL, and MD rack units, suggesting that these settings share the same degradation mechanism as the indoor tests. For the MD box unit, however, the degradation accelerates in the later stage, indicating that a different degradation mechanism may be emerging in the glass box environment.

The story for chemical change is slightly different. Under dry conditions, such as in AZ, the outdoor predictions align well with the observed data. For FL and the MD rack, the overall trend is captured but slightly over-predicted. This discrepancy may be due to the large extrapolated RH effect at high humidity levels and the effect of liquid water (which was not part of indoor experimental design), as indicated in Figure~\ref{fig:eff.fun.plot.chemical.change}(d). Since the indoor data include only two RH levels (0$\pect$ and 60$\pect$), applying the model to outdoor environments, where RH can frequently reach up to 100$\pect$, involves extrapolation.

For the MD box unit, a pattern similar to that of the yellowness index is observed, suggesting that further investigation of the glass box condition is warranted. Overall, the outdoor predictions are reasonably accurate, providing validation for the indoor acceleration models and offering insights into areas where both the models and experiments can be improved.

%%%%%%%%%%%%%%%%%%%%%%%%%%%%%%%%%%%%%%%%%%%%%%%%%%%%%%%%%%%%%%%%%%%%%%%%%%%%%%%%%%%%%%%%%%%%%%%%%%%
\section{Deep Learning Approach for Outdoor Predictions}\label{sec:deep.learning}
%%%%%%%%%%%%%%%%%%%%%%%%%%%%%%%%%%%%%%%%%%%%%%%%%%%%%%%%%%%%%%%%%%%%%%%%%%%%%%%%%%%%%%%%%%%%%%%%%%%
\subsection{Deep Learning Models}
In this section, we introduce a physics-informed deep learning (DL) model for degradation prediction. To motivate the model, we first consider the idea of parametric prediction. In the parametric setting, the degradation path is modeled as
\begin{align}\label{eqn:D(t)}
D(t)=\frac{A}{1+\exp\left\{-\dfrac{\log[s(t)]-\mu}{\sigma}\right\}},
\end{align}
for an increasing trend, where the effective dosage $s(t)$ is defined as in \eqref{eqn:outdoor.effective.dosage.cmpt}.
Note that in \eqref{eqn:outdoor.effective.dosage.cmpt}, parametric forms are used to model the effects of irradiance ($E(\lambda, \tau)$), temperature ($\TempK(\tau)$), and relative humidity ($\RH(\tau)$) at time $\tau$. Since DL models are highly flexible in capturing complex functional relationships, we extend the effective dosage model by representing it with a DL structure:
\begin{align}\label{eqn:DL.s(t)}
s(t)=\int_{0}^{t}o\left[E(\lambda, \tau), \TempK(\tau), \RH(\tau)\right]d\tau,
\end{align}
where $o\left[E(\lambda, \tau), \TempK(\tau), \RH(\tau)\right]$ is a function of irradiance, temperature, and RH, constructed using a DL model.

\begin{figure}
\begin{center}
\includegraphics[width=1.0\textwidth]{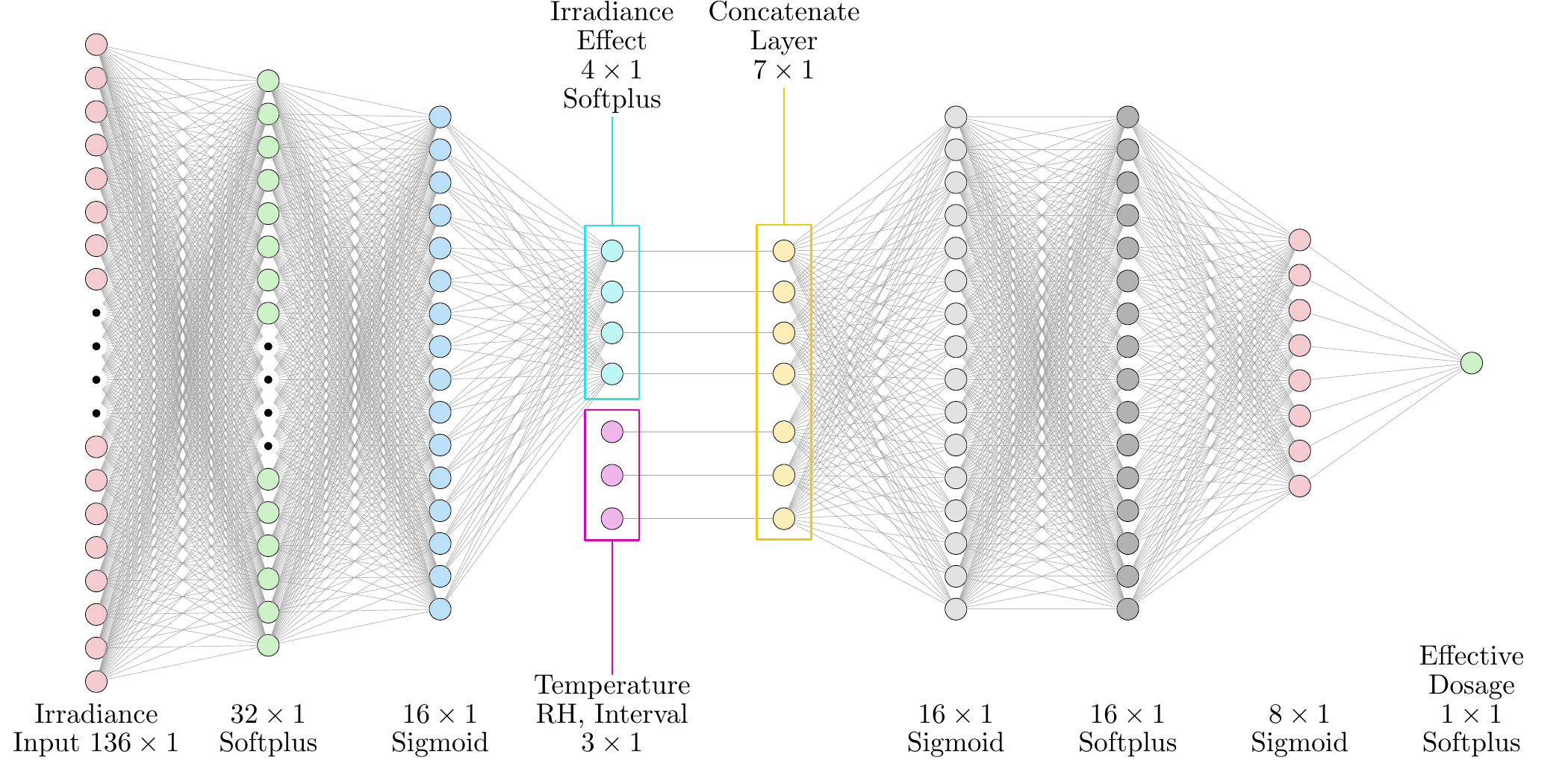}
\caption{Illustration of the neural network architecture for degradation prediction.}\label{fig:nn.structure}
\end{center}
\end{figure}

Figure~\ref{fig:nn.structure} illustrates the structure of the $o[,\cdot,\cdot,]$ function. In particular, the network consists of two branches. The first branch takes the irradiance at time $\tau$ as input, capturing the effects of spectrum and intensity. The second branch takes temperature, RH, and the interval length between two records as inputs. Its output is then concatenated with that of the first branch, as shown by the long red box in the middle of the figure. The combined input is subsequently processed through the network to produce $o[,\cdot,\cdot,]$ for time $\tau$.

To ensure monotonicity of the degradation path and allow extrapolation over time, we introduce a physics-informed DL model that builds on the general path function $g(s)$ derived from kinetics models. Specifically, the effective dosage $s(t)$ is substituted into \eqref{eqn:D(t)} to predict the degradation path, so the parameters $A$, $\mu$, and $\sigma$ still need to be estimated. Physical knowledge is incorporated through the function $g(s)$ to guide the overall degradation trend, while the DL component models the effects of covariates through $s(t)$, helping to address the limited sample size. The training of the DL model involves learning both $(A, \mu, \sigma)$ and the weight and bias parameters of the neural network. The model is implemented using PyTorch.

To train the model, we split the indoor data into a training set and a testing set. Since extrapolating covariates is challenging for a DL model, we include some outdoor units in the training set. Specifically, the ``FL'' and ``AZ'' units are used as the outdoor training set, while the ``MD rack'' and ``MD box'' units serve as the outdoor testing set. The indoor and outdoor training sets are then combined to form the final training set.

%%%%%%%%%%%%%%%%%%%%%%%%%%%%%%%%%%%%%%%%%%%%%%%%%%%%%%%%%%%%%%%%%%%%%%%%%%%%%%%%%%%%%%%
\subsection{Comparisons}
%%%%%%%%%%%%%%%%%%%%%%%%%%%%%%%%%%%%%%%%%%%%%%%%%%%%%%%%%%%%%%%%%%%%%%%%%%%%%%%%%%%%%%%

We compared three scenarios. The first uses the parametric model with the indoor training set to build the predictive model, referred to as ``Para. + Indoor.'' In this case, predictions for the outdoor setting can be made for all four units. The second uses the parametric model with the combined training set, referred to as ``Para. + Combined,'' where predictions are made only for the two outdoor testing units. The third uses the DL model with the combined training set, referred to as ``DL + Combined.''

\begin{table}[h]
\caption{Summary of sample sizes, reported as the number of units and the number of measurements, across different models and data settings for comparing statistical and DL predictions.}\label{tab:dl.sample.size}
\begin{center}
\begin{tabular}{c|c|cc|cc}\hline\hline
\multirow{2}{*}{Dataset}	& \multirow{2}{*}{Models}	& \multicolumn{2}{c|}{Indoor} &	\multicolumn{2}{c}{Outdoor}\\\cline{3-6}
	& 	& Training	& Prediction &	Training &	Prediction \\\hline
\multicolumn{2}{c|}{Sample Size}&\multicolumn{4}{c}{The Number of Units}\\\hline
           & Para. + Indoor     &98 & 14 & $-$ & 4 \\
Yellowness  & Para. + Combined  &98 & 14 & 2 & 2 \\
           & DL + Combined      &98 & 14 & 2 & 2 \\  \hline
  	       & Para. + Indoor	    &55 & 7 & $-$ & 4 \\
Chemical   & Para. + Combined   &55 & 7 & 2 & 2 \\
           & DL + Combined      &55 & 7 & 2 & 2 \\   \hline
\multicolumn{2}{c|}{Sample Size}&\multicolumn{4}{c}{The Number of Measurements}\\\hline
           & Para. + Indoor     &868 & 124 & $-$ & 39 \\
Yellowness  & Para. + Combined  &868 & 124 & 20 & 19 \\
           & DL + Combined      &868 & 124 & 20 & 19 \\  \hline
  	       & Para. + Indoor	    &552 & 70 & $-$ & 36 \\
Chemical   & Para. + Combined   &552 & 70 & 20 & 16 \\
           & DL + Combined      &552 & 70 & 20 & 16 \\   \hline \hline
\end{tabular}
\end{center}
\end{table}

We also considered two datasets: the yellowness index and the chemical change, consistent with the parametric models discussed in the previous sections. Table~\ref{tab:dl.sample.size} summarizes sample sizes, reported as the number of units and the number of measurements, across statistical models, DL models, and data settings. Table~\ref{tab:dl.comp.res} reports prediction accuracy measured by root mean square error (RMSE), mean absolute error (MAE), and $R^2$. For each dataset, the bolded values indicate the best value (i.e., the smallest value for RMSE and MAE, and the largest value for $R^2$) within the corresponding column for the dataset and criterion.

For the yellowness index, the DL model provides a better fit for both the indoor and outdoor training sets. In prediction, it also achieves the smallest RMSE for the indoor test set and the outdoor test set. Under MAE, the ``Para + Indoor'' setting performs best for the outdoor test set, which indicates that some large errors may have occurred, leading to a larger RMSE. The results based on $R^2$ are identical to those based on RMSE.
 For the chemical change data, however, the parametric model yields the smallest RMSE for both the indoor training and testing sets. The DL model attains the smallest RMSE for the outdoor training set, but its performance on the outdoor test set is substantially worse than that of the parametric model. The results under MAE and $R^2$ are the same as those based on RMSE.

These results highlight that while the DL model offers flexibility in capturing complex relationships and covariates, its ability to extrapolate to new covariate conditions is limited. Moreover, the outdoor dataset includes only four units, which constrains both model training and evaluation. Overall, the DL model can offer advantages in prediction, but the parametric model demonstrates greater robustness across datasets. In addition, the DL model sacrifices interpretability of model parameters compared with the parametric approach.

\begin{table}[h]
\caption{Comparisons of model performance across statistical models, DL models, and data settings. Performance is evaluated using RMSE, MAE, and $R^2$. The bolded numbers indicate the best value within each column for the corresponding dataset and criterion.}\label{tab:dl.comp.res}
\begin{center}
\begin{tabular}{c|c|cc|cc}\hline\hline
\multirow{2}{*}{Dataset}	& \multirow{2}{*}{Models}	& \multicolumn{2}{c|}{Indoor} &	\multicolumn{2}{c}{Outdoor}\\\cline{3-6}
	& 	& Training	& Prediction &	Training &	Prediction \\\hline
\multicolumn{2}{c|}{Criterion}&\multicolumn{4}{c}{RMSE}\\\hline
           & Para. + Indoor     &0.6163 & 0.7844 & $-$ & 1.1246 \\
Yellowness  & Para. + Combined  &0.6209 & 0.7429 & 0.7401 & 1.1639 \\
           & DL + Combined      &\textbf{0.4845} & \textbf{0.6967} & \textbf{0.4888} & \textbf{1.0889} \\   \hline
  	       & Para. + Indoor	    &\textbf{0.0527} & \textbf{0.0596} & $-$ & \textbf{0.1535} \\
Chemical   & Para. + Combined   &0.0531 & 0.0615 & 0.1090 & 0.1785 \\
           & DL + Combined      &0.0557 & 0.0755 & \textbf{0.0441} & 0.2100 \\    \hline
\multicolumn{2}{c|}{Criterion}&\multicolumn{4}{c}{MAE}\\\hline
           & Para. + Indoor     &0.4480 & 0.5837 & $-$ & \textbf{0.8003} \\
Yellowness  & Para. + Combined  &0.4497 & 0.5550 & 0.5975 & 0.8436 \\
           & DL + Combined      &\textbf{0.3351} & \textbf{0.5530} & \textbf{0.3050} & 0.8295 \\   \hline
  	       & Para. + Indoor	    &\textbf{0.0385} & \textbf{0.0418} & $-$ & \textbf{0.1203} \\
Chemical   & Para. + Combined   &0.0389 & 0.0432 & 0.0863 & 0.1356 \\
           & DL + Combined      &0.0411 & 0.0601 & \textbf{0.0350} & 0.1697 \\    \hline
\multicolumn{2}{c|}{Criterion}&\multicolumn{4}{c}{$R^2$}\\\hline
           & Para. + Indoor     &0.9505 & 0.8961 & $-$ & 0.5335 \\
Yellowness  & Para. + Combined  &0.9498 & 0.9068 & 0.7398 & 0.5853 \\
           & DL + Combined      &\textbf{0.9694} & \textbf{0.9180} & \textbf{0.8865} & \textbf{0.6370} \\   \hline
  	       & Para. + Indoor	    &\textbf{0.9426} & \textbf{0.8892} & $-$ & \textbf{0.2651} \\
Chemical   & Para. + Combined   &0.9417 & 0.8818 & 0.3810 & 0.3365 \\
           & DL + Combined      &0.9359 & 0.8224 & \textbf{0.8985} & 0.0815 \\    \hline \hline
\end{tabular}
\end{center}
\end{table}

%%%%%%%%%%%%%%%%%%%%%%%%%%%%%%%%%%%%%%%%%%%%%%%%%%%%%%%%%%%%%%%%%%%%%%%%%%%%%%%%%%%%%%%%%%%%%%%%%%%
\section{Concluding Remarks}\label{sec:conclusion}
%%%%%%%%%%%%%%%%%%%%%%%%%%%%%%%%%%%%%%%%%%%%%%%%%%%%%%%%%%%%%%%%%%%%%%%%%%%%%%%%%%%%%%%%%%%%%%%%%%%

In this paper, we develop both parametric and DL models to predict the degradation of polymeric components in PV systems, addressing an important challenge in assessing long-term performance. Degradation models play a central role in service life prediction, as they provide a quantitative framework for linking environmental stressors and material responses over time. A key element of this process is the construction of predictive models for the degradation path, which enables both understanding of underlying mechanisms and forecasting of future material reliability. The statistical (parametric) model provides a good fit and predictive performance for degradation paths and can be applied across datasets under various testing conditions. It also enables prediction for outdoor samples. While the yellowness index and chemical change data are used for illustration, the modeling framework can be extended to other degradation metrics, such as tensile strength.

The DL model provides considerable flexibility in capturing complex relationships among covariates and is capable of delivering accurate predictions under many conditions. Its strength lies in modeling nonlinear interactions that may be difficult to specify in a parametric form. However, when applied to outdoor prediction, parametric models that incorporate physical and chemical knowledge tend to perform more reliably. This robustness suggests that physics-informed parametric models remain an essential tool, particularly when extrapolation beyond the training data is required.

Although the training and test sets may appear small, they are the result of experiments conducted over multiple years and involve substantial effort and cost. A key aspect of this work is the integration of statistical and machine learning methods to enable prediction under such limited data conditions. At the same time, we acknowledge that the relatively small sample size is a limitation of the study, which may affect the generalizability of the results.

In contrast to \citet{Duanetal2017}, who employed a random effects model, this paper adopts nonlinear models (without random effects) to characterize degradation. The key difference lies in the objectives of prediction. Random effects models are well suited for capturing unit-to-unit variability when predicting the degradation of units that are part of the observed population. However, for predicting degradation paths of new outdoor units or making predictions at the population level, random effects models may not be appropriate. Nonlinear models, by directly linking degradation to underlying physical and environmental factors, provide a more suitable framework for these types of predictions.

Our methods can be extended to predict degradation for new materials. Since comprehensive statistical and DL models have already been developed for existing materials, updating only a subset of model parameters may be sufficient to generate predictions for the new materials. Consequently, a relatively small-scale data collection effort can be used to refine predictions by leveraging the established models. When new but similar materials are introduced, the existing statistical and DL frameworks can be adapted to provide predictions, with comparative experiments carried out to identify and quantify the differences between the old and new materials. In addition, statistical predictions of degradation can be extended to various geographical locations to account for the influence of different weather conditions. In terms of modeling, we considered the Arrhenius relationship for temperature, while the Eyring model (e.g., \citealt{EscobarMeeker2006}) could be explored in future work. Our prediction framework can also be applied to other degradation measurements, such as mechanical properties.

%%%%%%%%%%%%%%%%%%%%%%%%%%%%%%%%%%%%%%%%%%%%%%%%%%%%%%%%%%%%%%%%%%%%%%%%%%%%%%%%%%%%%%%%%
\section*{Supplementary Materials}

The following supplementary materials are available online.

\begin{description}
 
\item[Code and Data:] The data and computing code used in this paper are publicly available at: \url{https://doi.org/10.5281/zenodo.19981142}.

\end{description}

\section*{Disclaimer}
Certain commercial products or equipment are described in this paper to specify adequately the experimental procedure. In no case does such identification imply recommendation or endorsement by the National Institute of Standards and Technology, nor does it imply that it is necessarily the best available for the purpose.

%%%%%%%%%%%%%%%%%%%%%%%%%%%%%%%%%%%%%%%%%%%%%%%%%%%%%%%%%%%%%%%%%%%%%%%%%%%%%%%%%%%%%%%%%%%%%%%%%
\bibliographystyle{chicago}

%\bibliography{ref}

%%%%%%%%%%%%%%%%%%%%%%%%%%%%%%%%%%%%%%%%%%%%%%%%%%%%%%%%%%%%%%%%%%%%%%%%%%%%%%%%%%%%%%%
\end{document}